\documentclass[10pt,journal,comsoc]{IEEEtran}
\usepackage[T1]{fontenc}
\usepackage[latin9]{inputenc}
\usepackage{float}
\usepackage{units}
\usepackage{amsmath}
\usepackage{amsthm}
\usepackage{amssymb}
\usepackage{graphicx}

\usepackage{xurl}

\usepackage[unicode=true,
 bookmarks=true,bookmarksnumbered=true,bookmarksopen=true,bookmarksopenlevel=1,
 breaklinks=false,pdfborder={0 0 0},pdfborderstyle={},backref=false,colorlinks=false]
 {hyperref}
\hypersetup{pdftitle={Your Title},
 pdfauthor={Your Name},
 pdfpagelayout=OneColumn, pdfnewwindow=true, pdfstartview=XYZ, plainpages=false}

\makeatletter

\providecommand{\tabularnewline}{\\}
\floatstyle{ruled}
\newfloat{algorithm}{tbp}{loa}
\providecommand{\algorithmname}{Algorithm}
\floatname{algorithm}{\protect\algorithmname}

\theoremstyle{plain}
\newtheorem{thm}{\protect\theoremname}
\theoremstyle{plain}
\newtheorem{prop}{\protect\propositionname}

\usepackage[caption=false,font=normalsize,labelfont=sf,textfont=sf]{subfig}

\@ifundefined{showcaptionsetup}{}{%
 \PassOptionsToPackage{caption=false}{subfig}}
\usepackage{subfig}
\makeatother

\providecommand{\propositionname}{Proposition}
\providecommand{\theoremname}{Theorem}

\begin{document}
\title{Dual Auction Mechanism for Transaction Relay and Validation in
Complex Wireless Blockchain Network}
\author{Weiyi~Wang, Yutao~Jiao, Jin~Chen, Wenting~Dai, Jiawen Kang~and~Yuhua~Xu
\thanks{Weiyi Wang, Yutao Jiao, Jin Chen, and Yuhua
Xu are with the College of Communications Engineering, Army Engineering University of PLA, Nanjing 210007, China, E-mail: \href{http://ytjiao@aeu.edu.cn}{ytjiao@aeu.edu.cn}.}
\thanks{Wenting Dai is with the School of Communications and Information Engineering, Nanjing University of Posts and Telecommunications, Nanjing 210003, China.}
\thanks{Jiawen Kang is with the School of Automation, Guangdong University of Technology, Guangzhou 510006, China, and also with the Key Laboratory of Intelligent Information Processing and System Integration of IoT, Ministry of Education, Guangzhou 510006, China.}
\thanks{(Corresponding author: Yutao Jiao and Jin Chen)}}

\IEEEtitleabstractindextext{
\begin{abstract}
In traditional public blockchain networks, transaction fees are only allocated to full nodes (i.e., miners). However, the lack of relay rewards reduces the willingness of light nodes to relay transactions, especially in the energy-constrained complex wireless network. This paper proposes a novel dual auction mechanism to allocate transaction fees for relay and validation behaviors in the wireless blockchain network. The dual auction mechanism consists of two sub-auction stages: the relay sub-auction and the validation sub-auction. In the relay sub-auction, relay nodes choose transactions based rewards to forward. Besides, relay nodes adjust the relaying probability through a no-regret algorithm to improve efficiency. In the validation sub-auction, full nodes select transactions using Vickrey-Clarke-Grove (VCG) mechanism to construct the block. We prove that the designed dual auction mechanism is Incentive Compatibility (IC), Individual Rationality (IR), and Computational Efficiency (CE). We also derive the upper bound of the social welfare difference between the social optimal auction and our proposed one. Extensive simulation results demonstrate that the proposed dual auction mechanism decreases energy and bandwidth resource consumption and effectively improves social welfare without sacrificing the throughput and the security of the wireless blockchain network.
\end{abstract}


}
\maketitle

\IEEEdisplaynontitleabstractindextext{}

\IEEEpeerreviewmaketitle{}

\section{Introduction}

\IEEEPARstart{B}{lockchain} has been a hot topic in industry and academia due to its distributed, decentralized, and truthful characteristics \cite{zheng2018blockchain}. Currently, most blockchain applications rely on desktop terminals, which hinders wider adoption and implementation of blockchain technology. Fortunately, integrating blockchain and wireless communication eliminates the limitations of time and space, which attracts more users to participate in the construction of blockchain. For instance, integrating blockchain and Mobile Ad Hoc Network (MANET) not only provides a truthful platform for mobile data sharing but also enriches blockchain applications. 

However, there are several challenges to applying existing blockchain technologies directly to complex wireless networks. One intractable problem is that the traditional incentive mechanisms of blockchain are not appropriate for wireless nodes. Taking the Proof-of-Work (PoW) consensus as an example (e.g., Bitcoin \cite{nakamoto2008peer}, Ethereum \cite{buterin2017ethereum}), full nodes are responsible for solving the hash  puzzle and proposing blocks, and light nodes help relay transactions (Txs) and blocks. The consensus mechanism mainly focuses on the nodes' computing capacities, and the transaction fee is paid solely to full nodes. It is unfair for the light nodes since they consume energy and bandwidth resources to relay the transactions and blocks. Lacking relay rewards decreases the impetus of light nodes to help others, which increase the delay of messages and compromises the robustness and throughput of the blockchain network. 

Furthermore, traditional blockchain networks use the broadcast protocol to spread information. The continual communication behaviors consume vast energy and bandwidth resources, which contradicts wireless nodes' constrained energy and bandwidth resources. Various types of wireless nodes also bring challenges in message propagation. Heterogeneous nodes connect each other in an irregular mode, which leads to the complex wireless network topology (e.g. Erdos-Renyi (ER) network, Barabasi-Albert (BA) network, and Watts-Strogatz (WS) network). The complex wireless blockchain network introduces additional communication overhead and further reduces the efficiency during the existing relay mechanism. The transaction relay accounts for a higher proportion in throughput and security of blockchain. Thus, the incentive mechanism should be improved to adapt to the complex wireless blockchain network.

According to above analysis, both relay and validation behaviors should be rewarded in the wireless blockchain network. Due to the introduction of the relay reward, the mechanism should resist Sybil attacks \cite{babaioff2012bitcoin}, which means nodes cannot forge fake relay behaviors to obtain profits. Besides, the nodes are guided to relay transactions under the incentive mechanism to avoid the waste of energy and bandwidth. Considering the reward allocation and the complex wireless environment, the designed incentive mechanism aims to have the following properties: 
\begin{enumerate}
\item The transaction fee is allocated for both relay and validation behaviors. 
\item The incentive mechanism can guide relay nodes to adjust propagation strategies adaptively and forward transactions honestly.
\item The incentive mechanism should not decrease the performance of throughput and compromise the security of the wireless blockchain system. 
\end{enumerate}

Based on the above requirements, we propose a novel \textbf {dual auction mechanism} for the wireless blockchain network. When nodes produce transactions, they add the transaction fees as bids and broadcast them on the network. The dual auction mechanism includes two sub-auctions. In the first sub-auction, the relay nodes receive the transaction and decide whether forward the transaction according to the bid and relaying probability. In the second sub-auction, the full nodes select transactions from the transaction pool to pack the block based on bids. Transaction fees are paid to relay and validation nodes once the block containing corresponding transactions is verified and confirmed. Therefore, in the designed dual auction mechanism, transaction producers bidding transaction fees act as bidders, other light and full nodes contributing communication and computing resources are as sellers, and the blockchain functions as the auctioneer. The significant contribution of this paper can be summarized as follows:
\begin{itemize}
\item We propose a dual auction mechanism including two sub-auctions to split the transaction fee into relay and validation rewards. In the dual auction mechanism, the wireless blockchain system not only allocates rewards for full nodes but also compensates relay nodes for energy and bandwidth consumption, encouraging more nodes to contribute their resources to the blockchain.
\item We elaborately construct the reward allocation mechanism while considering the Sybil attack during the relay phase. By preventing Sybil attacks, our mechanism ensures the integrity and safety of the wireless blockchain system.
\item To reduce the bandwidth occupancy, we use the no-regret algorithm to adjust the relaying probability of nodes in the first sub-auction. The relay nodes prefer to assist transaction producers whose transactions are more likely to be confirmed on the blockchain by adaptively altering the relaying probability.
\item In the designed dual auction mechanism, we prove that the whole dual auction is Incentive Compatibility (IC), Individual Rationality (IR), and Computational Efficiency (CE). Especially, we theoretically analyze the upper bound of the social welfare difference between the devised and the optimal social mechanisms to show the effectiveness of our mechanism.
\item Simulation results show that the proposed mechanism decreases energy and bandwidth consumption without compromising the Transactions per Second (TPS), and adapts to the network with node failures and Byzantine attacks in wireless environments.
\end{itemize}

The rest of this paper is organized as follows. Section 2 reviews the related work. Section 3 demonstrates the wireless blockchain system. The proposed dual auction mechanism is discussed in Section 4. In Section 5, we give the theoretical proof of the property of the dual auction mechanism. Simulation results of the performance of the designed mechanism are presented in Section 6. Finally, Section 7 concludes the paper. Table 1 lists the main notations of this paper.

\begin{table}[tbh]
\caption{Main Notations}
\centering{}%
\begin{tabular}{ll}
\hline 
Notation & Definition\tabularnewline
\hline 
$\mathcal{B}$ & Blockchain\tabularnewline
$\mathcal{P}$ & Transaction pool\tabularnewline
$\mathcal{L}$ & List of \emph{Tx\_msg}s\tabularnewline
$X$ & Number of light nodes\tabularnewline
$Y$ & Number of full nodes\tabularnewline
$M$ & Maximum number of transactions in a block\tabularnewline
$N$ & Relaying window\tabularnewline
$F$ & Transaction fee\tabularnewline
$V$ & Transaction value\tabularnewline
$T_{c}$ & Expected delay of confirmation\tabularnewline
$T_{s}$ & Timestamp\tabularnewline
$L$ & Maximum length of relay path\tabularnewline
$l_{t}$ & Remaining times of relay\tabularnewline
$n_{f}$ & Number of selected neighbors to forward\tabularnewline
$p$ & Relaying probability\tabularnewline
$\delta$ & Indicator variable of transaction confirmation\tabularnewline
$\eta$ & Indicator variable of relay message confirmation\tabularnewline
$\sigma$ & Indicator variable of block confirmation\tabularnewline
$v$ & Reward\tabularnewline
$c_{f}$ & Relay cost\tabularnewline
$c_{v}$ & Validation cost\tabularnewline
$\alpha$ & Constant coefficient of transaction fee allocation\tabularnewline
$\gamma$ & Learning rate\tabularnewline
$r$ & Length of relay path\tabularnewline
$w_{a}$ & Weight of action\tabularnewline
$w_{t}$ & Weight of time\tabularnewline
$b$ & Bid\tabularnewline
$\pi$ & Payment \tabularnewline
$\omega$ & Winner\tabularnewline
\hline 
\end{tabular}
\end{table}

\section{Related Work}

To expand the blockchain application, most researchers focus on integrating the blockchain and wireless networks \cite{reyna2018blockchain}. For example, the authors utilized blockchain to assist data aggregation, which reduced the communication and computing overhead for wireless sensors \cite{lu2021edge}. However, nodes connect in wireless links in the wireless network, and the communication capacity has a more decisive impact on blockchain performance. In \cite{decker2013information}, the authors analyzed the propagation delay of different types of messages in Bitcoin and deduced the expression between the propagation delay and the stale rate of blocks. The authors in \cite{shahsavari2020theoretical} introduced different propagation protocols in Bitcoin and concluded that the block size, network size, node connectivity, and proportion of relay nodes affect blockchain performance.  Signal to Interference plus Noise Ratio (SINR) influences the success rate of transactions and the efficiency of consensus algorithms \cite{xu2020raft,onireti2019viable}.  In \cite{wei2020rethinking}, the authors derived that the malicious node with less than 50\% computing power could achieve a double-spending attack through the communication advantage. 

Recently, some research has focused on the significance of message relay in the blockchain network. In \cite{Hao2019BlockP2PEF}, \cite{Rohrer2019KadcastAS}, \cite{Zhu2022DesignOL}, authors utilized structured methods to reduce the delivery latency. Selecting neighbors to relay is investigated for shortening the propagation time \cite{Aoki2019ProximityNS}, \cite{Mao2020PerigeeEP}. In \cite{Qiu2023AGP}, the authors  proposed a geography-based P2P transmission protocol and analyzed the traffic pattern. The authors in \cite{Naumenko2019ErlayET} designed a relay protocol, Erlay, aiming to reduce the bandwidth consumption for Bitcoin.
In \cite{Zhou2023MercuryFT}, the authors propose a protocol combining the geography and cluster to reduce the transaction latency. 

Our paper mainly focuses on improving the incentive mechanism in wireless blockchain systems. The traditional transaction fee mechanisms in Bitcoin \cite{nakamoto2008peer} and Ethereum \cite{buterin2017ethereum} are believed to be the Generalized First Price (GFP) auction. In EIP-1559 \cite{roughgarden2020transaction}, the transaction fee mechanism is improved but still a GFP auction, which leads nodes to pay more fees for transaction confirmation. Besides, researchers discussed the potential path for relay icentivization \cite{Eth2024Relay}. In \cite{xue2021incentive}, the authors investigated the rewards for full nodes considering the cost and strategy in the mining pool. The authors in \cite{wang2021multi} and \cite{wang2022connectivity} use the contract theory to incentivize nodes considering the connectivity. In \cite{li2020novel}, the authors proposed a Generalized Second Price (GSP) auction mechanism in blockchain but did not consider the relay reward. In \cite{machado2021blockchain} and \cite{ling2021data}, the authors investigated the relay reward to encourage nodes to route data. Nevertheless, they considered using blockchain as a ledger to record the relay information and did not improve the incentive mechanism for blockchain. The authors in \cite{kang2018incentivizing} proposed an incentive mechanism to accelerate block propagation. The mechanism offered the transaction fee for transaction verification of full nodes but did not award the relay behaviors. The most closely related papers are \cite{babaioff2012bitcoin} and \cite{ersoy2018transaction}. In \cite{babaioff2012bitcoin}, the authors proposed a Sybil-proof mechanism to incentivize information propagation under a forest network of $t$ complete $d$-ary trees. In \cite{ersoy2018transaction}, the authors studied the relay reward and proposed a transaction fee mechanism to allocate the fee to relay nodes. However, the mechanism cannot resist the Sybil attack with the 1-connected network. Besides, neither \cite{babaioff2012bitcoin} nor \cite{ersoy2018transaction} fully considered the wireless environments and communication factors.

\begin{figure*}[t]
 \includegraphics[scale=0.5]{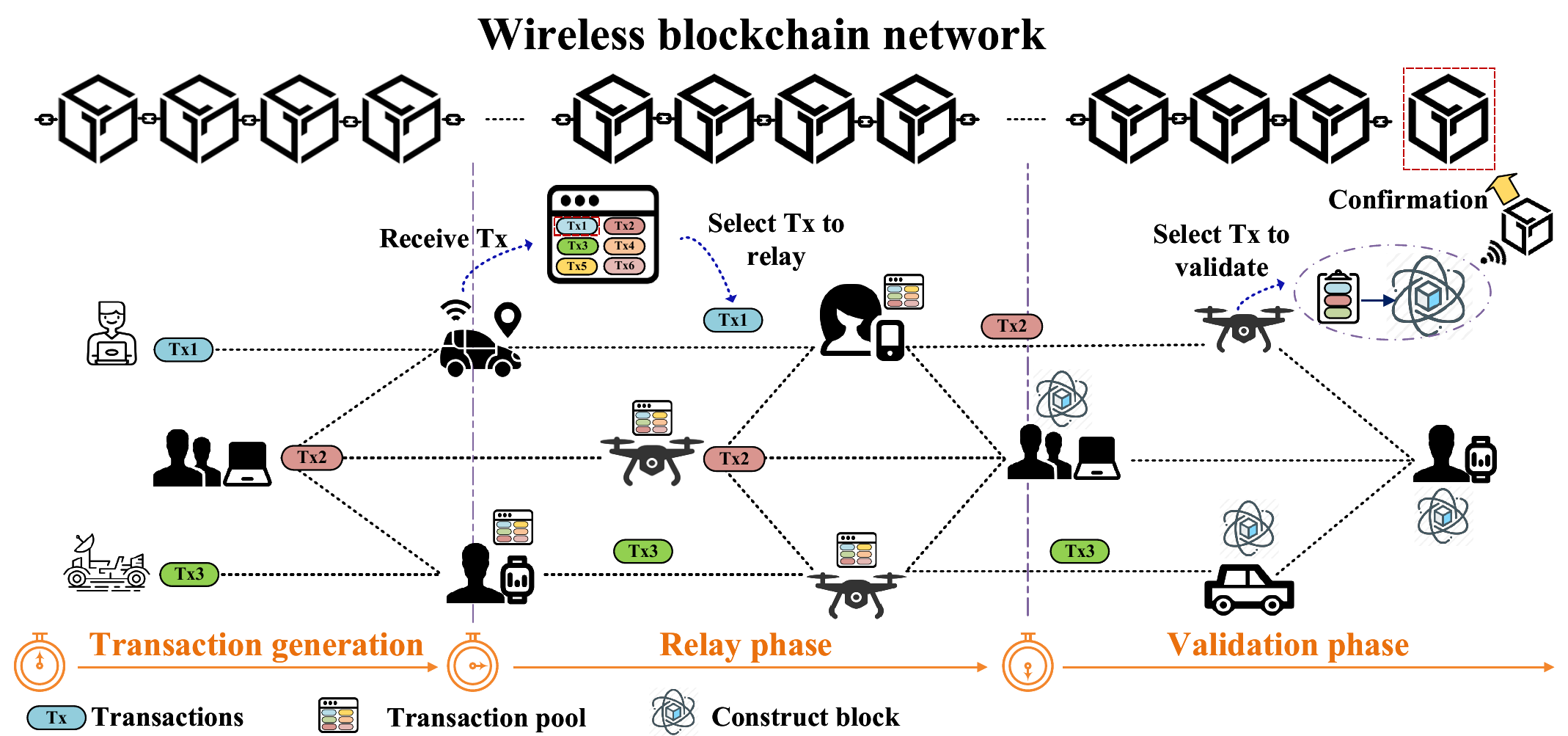}\caption{The relay and validation process in the complex wireless blockchain network.}
\end{figure*}

\section{System Model: Wireless Blockchain Network}

Fig. 1 illustrates the wireless blockchain network. During the propagation process of transactions, relay nodes forward transactions to full nodes from the transaction pool, and then full nodes select transactions to construct the block. After validating and confirming the new block, relevant light and full nodes will receive the relay and validation rewards.

\subsection{Wireless Blockchain System}

We consider a wireless blockchain system without fixed network infrastructures. Every node connects in wireless links and is free to join or leave the blockchain network. Due to the complex wireless environment and heterogeneity of nodes, we assume that the nodes have the probability of being at fault and keep silent during the fault time. Similar to Bitcoin \cite{nakamoto2008peer} and Ethereum \cite{buterin2017ethereum}, we adopt the PoW-based algorithm as the consensus, in which nodes compete for the proprietorship of the block by solving the hash puzzle. Besides, only the block on the longest chain is legal. Otherwise, it is invalid. 

There are $X+Y$ nodes in the wireless blockchain system, which are divided into two categories: $X$ \emph{light nodes} and $Y$ \emph{full nodes}. Full nodes are responsible for solving the hash puzzle (i.e., mining) and relaying messages. Light nodes just afford to relay messages. All nodes are assumed to be selfish and only choose behaviors that bring non-negative utility. 

The wireless blockchain network mainly uses three message types: \emph{Tx\_msg}, \emph{Block\_msg}, and \emph{Syn\_msg}. \emph{Tx\_msg} represents the transaction produced by light nodes and is expected to reach the full nodes.\emph{ Block\_msg} is the block proposed by full nodes, which is larger than \emph{Tx\_msg} (e.g., in Bitcoin, the average size of \emph{Tx\_msg} is 250 Bytes, but \emph{Block\_msg} is 500,000 Bytes) and expected to spread across the whole network. The \emph{Tx\_msg} is deemed to be confirmed and valid if it is contained in the \emph{Block\_msg}. Since the block is related to both light and full nodes, we assume that all nodes relay the \emph{Block\_msg} spontaneously and a \emph{Block\_msg} at most includes $M$ \emph{Tx\_msg}s. \emph{Syn\_msg} is used to help synchronize the block. At a fixed interval, the node broadcasts the \emph{Syn\_msg} to its neighbors, and then neighbors send their latest block height to the node. The node will request and synchronize the blockchain the neighbor stores when the
received block height is $n_{c}$ larger than itself. For example, $n_{c}$ is set to 6 in the Bitcoin system.

We use \emph{gossip} protocol in Bitcoin \cite{nakamoto2008peer} as the broadcast mode. Therefore, the message has the form $\left\langle l_{t},n_{f}\right\rangle $, where $l_{t}$ represents the Time To Live (TTL) (i.e., the remaining relay times), and $n_{f}$ represents the fanout (i.e., the number of selected neighbors). We set $L$ as the maximum value of $l_{t}$. After each relay, the $l_{t}$ of the message decreases by 1. Once $l_{t}$ becomes 0, the message will not be relayed.

\subsection{Transaction Relay and Validation Mechanism for Wireless Blockchain}

Traditional blockchain systems prefer to allocate all transaction fees to full nodes because they consume vast resources to propose blocks. Nevertheless, this reward allocation mechanism is incompatible with wireless networks, where the communication factors are crucial in confirming transactions and blocks. As mentioned in \cite{9155451}, the communication advantage influences the propagation delay, which determines the winning probability during the fork competition. Wireless blockchain designers must attach importance to nodes' communication contributions and distribute the transaction fee to light nodes. Thus, the transaction fee $F$ comprises the following parts:

\begin{equation}
F=F_{v}+F_{f},\label{eq:Transcation fee}
\end{equation}
where $F_{v}$ is the validation reward, $F_{f}$ is the relay reward, and $F_{v}\geq F_{f}$.

Note that though the relay process is essential for wireless blockchain networks, the block is produced by full nodes ultimately, and then transactions are confirmed. It is why the validation rewards $F_{v}$ should be no less than the relay rewards $F_{f}$ so that full nodes still pay attention to verify, which is vital for the security of the wireless blockchain system. Moreover, light nodes are willing to help relay \emph{Tx\_msg} under this incentive mechanism, especially in wireless networks. Therefore, the improved mechanism can encourage more nodes to contribute resources to the wireless blockchain.

\subsection{Problem Formulation}

As mentioned above, light nodes cannot propose the block, and they just obtain the relay reward. On the one hand, the light node produces transactions and pays the corresponding fee. On the other hand, the light node can relay others' transactions to earn the transaction fees. Considering the limited resources of nodes in the wireless blockchain network, light nodes are not required to relay every received \emph{Tx\_msg}. According to different source nodes, light nodes have different relaying probability $p$ and adjust the probability adaptively based on the results of transaction confirmation. The utility function of light node $i$ is defined as follows:

\begin{equation}
U_{f}^{i}=\sum_{k=1}^{K}\delta_{k}^{i}\left(\mathop{V_{k}^{i}-F_{k}^{i}}\right)+\sum_{j=1}^{J}\mathop{\left(\eta_{ij}F_{f}^{j}-p_{ij}c_{f}^{i}\right)},\label{eq:Peer utility function}
\end{equation}
where $V_{k}^{i}$ is the valuation of the $k$-th transaction of node $i$, $F_{k}^{i}$ is the paid transaction fee of the $k$-th transaction of node $i$, $p_{ij}$ is the relaying probability that node $i$ relays the transaction from node $j$, $c_{f}^{i}$ is the cost of node $i$ for relaying a transaction, $\delta_{k}^{i},\eta_{ij}\in\{0,1\}$ are the indicator variable. $\delta_{k}^{i}=1$ means the $k$-th transaction of node $i$ is successfully confirmed and on the blockchain, and $\eta_{ij}=1$ indicates the transaction of node $j$ relayed by node $i$ is confirmed. Thus, the first term of equation \eqref{eq:Peer utility function} represents the utility of producing transactions, and the second is the utility of relaying transactions.

From the perspective of transaction propagation, full nodes are the destination of \emph{Tx\_msg}. Besides, each full node is able to relay and validate transactions. The relay rewards obtained by full nodes is the same as equation \eqref{eq:Peer utility function}. We mainly focus on the validation rewards of full nodes in equation \eqref{eq:miner utility function}. The full node will fill the block with as many transactions as possible to maximize the utility. We assume that every full node puts $N$ \emph{Tx\_msg}s into \emph{Block\_msg} to obtain more transaction fees. The utility function of full node $l$ is given in the following:

\begin{equation}
U_{v}^{l}= \sum_{j=1}^{J}\mathop{\left(\eta_{lj}F_{f}^{j}-p_{lj}c_{f}^{l}\right)} + \sum_{s=1}^{S}\left(\sigma_{s}^{l}\sum_{n}F_{v}^{n}-c_{v}^{l}\right),\label{eq:miner utility function}
\end{equation}
where $c_{v}^{l}$ is the cost of full node $l$, and $\sigma_{s}^{l}\in\left\{ 0,1\right\}$ is the indicator variable. $\sigma_{s}^{l}=1$ represents the proposed $s$ th block on the longest legal chain and vice versa. Note that block generation is a stochastic process, each full node generates the block based on the hash power. However, even if the block is generated, it may not be on the longest legal chain due to the forking competition. We use the indicator $\sigma_{s}^{l}$ to indicate whether the block is accepted by the majority of nodes.

Based on equations \eqref{eq:Peer utility function} and \eqref{eq:miner utility function}, we formalize the problem of social welfare maximization as follows:

\begin{align}
\max_{\boldsymbol{p},\boldsymbol{F}} & \sum_{i=1}^{X}U_{f}^{i}+\sum_{l=1}^{Y}U_{v}^{l},\label{eq:social welfare}\\
s.t.\nonumber \\
 & \delta_{k}^{i},\eta_{ij},\sigma_{s}^{l}\in\left\{ 0,1\right\} ,\\
 & 0\leq p_{ij}\leq1,\\
 & F_{k}^{i}\leq V_{k}^{i}.\label{eq:tx fee cons}
\end{align}

In equation \eqref{eq:social welfare}, our objective is to determine the relaying probability and transaction fee to maximize the social welfare, inequation \eqref{eq:tx fee cons} represents that the transaction fee cannot exceed the transaction value. 

\begin{figure}[t]
    \includegraphics[scale=0.5]{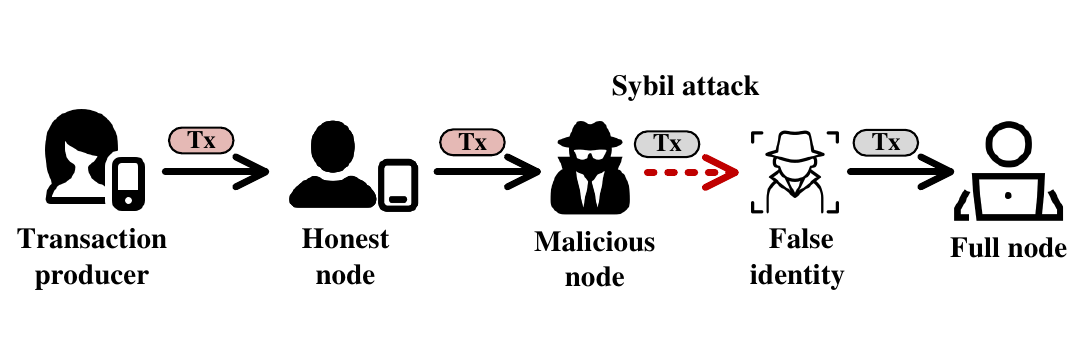}
    \caption{Sybil attack mode in the relay process.}
\end{figure}

\subsection{Adversary Model}

We mainly consider two common adversary model types: \emph{Sybil Attack} and \emph{Double Spending Attack}.
\begin{enumerate}
\item \emph{Sybil Attack}. Since the incentive mechanism includes the relay rewards, malicious nodes may launch Sybil attacks by forging the relay record to earn extra transaction fees. Both light and full nodes can launch Sybil attacks. As shown in Fig.2, the malicious node signs multi fake identities on the \emph{Tx\_msg} while relaying it. Full nodes can append the fake identities on the relay list and then put the \emph{Tx\_msg} into \emph{Block\_msg.} Due to the Sybil attack, the malicious nodes obtain undeserved utility, which undermines the security and fairness of the wireless blockchain system. 
\item \emph{Double Spending Attack}. This adversary model is the classical attack in blockchain systems, and only full nodes can start the double spending attack. Malicious full nodes produce a \emph{Block\_msg} composing the \emph{Tx\_msg} that contradicts the previous block. Thus, the blockchain will fork at the height of the contradictory blocks. Once the blockchain, including the \emph{Block\_msg} produced by malicious full nodes, becomes the longest legal chain, the double spending attack is successful. The double spending attack makes forks obtain utilities, which causes severe damage to the immutability of blockchain systems. 
\end{enumerate}

In a nutshell, we should design a mechanism that allocates transaction fees fairly while reducing resource consumption and maintaining security in the wireless blockchain network.

\begin{figure*}[t]
    \includegraphics[scale=0.88]{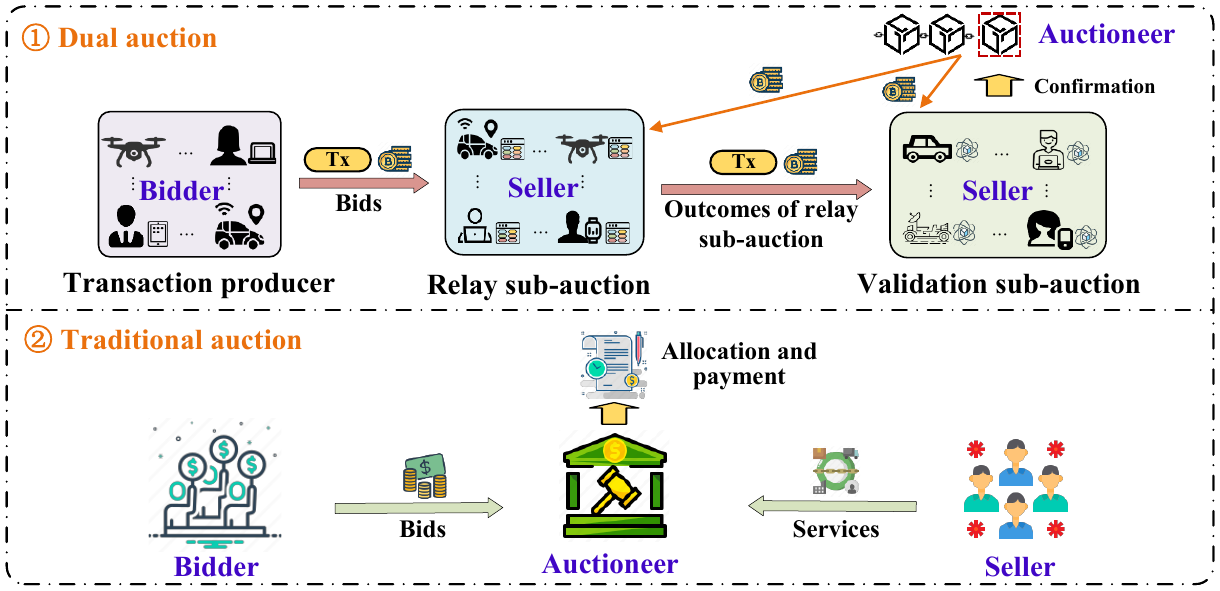}
    \caption{Comparision of the proposed dual auction and traditional auction.}
\end{figure*}

\section{Dual Auction Mechanism for Wireless Blockchain Network}

In this section, we design a dual auction mechanism comprising two sub-auctions: the relay sub-auction and the validation sub-auction. In the first stage, relay nodes obtain transaction fees through routing \emph{Tx\_msg}s. In the second stage, full nodes verify and propose blocks to earn rewards. 

Note that our dual auction is different from the traditional auction. As shown in Fig. 3, the dual auction consists of two sub-auctions for covering the relay and validation of the \emph{Tx\_msg}. We integrate the two sub-auctions into the \emph{Tx\_msg} propagation process. Producers just bid once for completing two sub-auctions. In the validation sub-auction, the generation and confirmation of blocks determine the success of both sub-auctions. However, the relay sub-auction determines whether full nodes could receive \emph{Tx\_msg}s. The whole dual auction fits the protocol operation of blockchain and simplifies the auction procedure, which helps blockchain to apply in the wireless network.

\subsection{Sybil-proof Incentive Mechanism Design}

Defending against the Sybil attack is the prerequisite to guaranteeing the effectiveness and fairness of the designed incentive mechanism. Since the transaction fee $F$ is determined while the \emph{Tx\_msg} is produced, the point is the allocation of the transaction fee. Thus, malicious nodes launch Sybil attacks to obtain more transaction fees. From the economics perspective, the incentive mechanism is Sybil-proof if it satisfies the following conditions:

\begin{equation}
F_{r}^{g}\geq F_{r+u}^{g}+F_{r+u}^{g+1}+\cdots+F_{r+u}^{g+u},\label{eq:Reward allocation}
\end{equation}
where $F_{r}^{g}$ is the relay reward of the $g$-th node with the $r$ length path, and $g=r$ is the full nodes' reward. From the equation \eqref{eq:Reward allocation}, the longer the relay path is, the fewer transaction fees the node obtains with the same relay sequence. Precisely, $F_{r}^{g}\geq F_{r+u}^{g}$. Besides, the additional amount of relay nodes cannot increase the total transaction fee, which means $\sum_{g=1}^{r}F_{r}^{g}\geq\sum_{g=1}^{r+u}F_{r}^{g}$. Thus, equation \eqref{eq:Reward allocation} ensures that the malicious node cannot obtain extra profit through launching Sybil attacks. The \emph{Tx\_msg }producer need not pay fees with the increase of relay nodes.

For design convenience, we give Theorem 1 as follows:
\begin{thm}
The equation \eqref{eq:Reward allocation} holds if and only if $F_{r}^{g}\geq F_{r+1}^{g}+F_{r+1}^{g+1}$ holds.
\end{thm}
\begin{IEEEproof}
With the equation $F_{r}^{g}\geq F_{r+u}^{g}$ and equation $F_{r}^{g}\geq F_{r+1}^{g}+F_{r+1}^{g+1}$, we can derive

\begin{align}
F_{r}^{g} & \geq F_{r+1}^{g}+F_{r+1}^{g+1}\geq F_{r+1}^{g}+F_{r+2}^{g+1}+F_{r+2}^{g+2}\nonumber \\
 & \geq\ldots\geq F_{r+1}^{g}+F_{r+2}^{g+1}+\ldots+F_{r+u}^{g+u}\nonumber \\
 & \geq F_{r+u}^{g}+F_{r+u}^{g+1}+\ldots+F_{r+u}^{g+u}.
\end{align}
\end{IEEEproof}
Theorem 1 implies that if the incentive mechanism satisfies $F_{r}^{g}\geq F_{r+1}^{g}+F_{r+1}^{g+1}$, the mechanism is Sybil-proof. Based on Theorem 1, we give the designed incentive mechanism in thefollowing:

\begin{equation}
F_{r}^{g}=\begin{cases}
(\frac{1}{1+\alpha})^{r-1}\alpha^{g}F, & g=1,\ldots,r-1,\\
(\frac{1}{1+\alpha})^{r-1}F, & g=r,
\end{cases}\label{eq:designed allocation mechanism}
\end{equation}
where $0\leq\alpha\leq1$ is the constant coefficient. The first term is the relay reward, and the second is the validation reward. Note that there is no relay reward, and the transaction fee is all allocated to the full node when $r=1$. Besides, light nodes at the front of the relay sequence are incentivized more due to their higher importance.

\begin{prop}
The dual auction mechanism is effective and Sybil-proof.
\end{prop}
\begin{IEEEproof}
First, we prove that full nodes cannot obtain extra rewards by launching Sybil attacks. According to the equation \eqref{eq:designed allocation mechanism}, the full node's reward is

\begin{equation}
R_{v}=(\frac{1}{1+\alpha})^{r-1}F.
\end{equation}

If the full node makes a Sybil attack, its allocated fee is

\begin{equation}
R_{v}^{'}=(\frac{1}{1+\alpha})^{r}F+(\frac{1}{1+\alpha})^{r}\alpha^{r}F.
\end{equation}

Then we can derive 

\begin{align}
R_{v}-R_{v}^{'} & =\left(1-\frac{1+\alpha^{r}}{1+\alpha}\right)(\frac{1}{1+\alpha})^{r-1}F\nonumber \\
 & =(\frac{\alpha-\alpha^{r}}{1+\alpha})(\frac{1}{1+\alpha})^{r-1}F\nonumber \\
 & \geq0.
\end{align}

Thus, we prove that the mechanism is Sybil-proof for full nodes. Next, we prove it is also Sybil-proof for light nodes. The reward of the honest light node is

\begin{equation}
R_{f}=(\frac{1}{1+\alpha})^{r-1}\alpha^{g}F.\label{eq:honest peer}
\end{equation}

If the light node chooses to launch a Sybil attack, the reward will be

\begin{equation}
R_{f}^{'}=(\frac{1}{1+\alpha})^{r}\alpha^{g}F+(\frac{1}{1+\alpha})^{r}\alpha^{g+1}F.\label{eq:malicious peer}
\end{equation}

Similarly, we can obtain

\begin{equation}
R_{f}-R_{f}^{'}=\left(1-\frac{1+\alpha}{1+\alpha}\right)(\frac{1}{1+\alpha})^{r-1}\alpha^{g}F=0.
\end{equation}

Note that the limit of $l_{t}$, if the light node launches a Sybil attack, the effective $l_{t}$ will change into $l_{t}-1$. It decreases the probability of the \emph{Tx\_msg} reaching full nodes. Meanwhile, the Sybil attack reduces rewards for other nodes, and others are unwilling to relay the \emph{Tx\_msg}. Therefore, the Sybil-proof of the mechanism is proved.

Last, we prove the effectiveness of the allocation mechanism. The total reward allocated to nodes is

\begin{align}
\sum_{g=1}^{r}R^{g} & =(\frac{1}{1+\alpha})^{r-1}F+\sum_{g=1}^{r-1}(\frac{1}{1+\alpha})^{r-1}\alpha^{g}F\nonumber \\
 & =\frac{1-\alpha^{r}}{(1-\alpha)(1+\alpha)^{r-1}}F\nonumber \\
 & =\frac{(1-\alpha)(1+\alpha+\ldots\alpha^{r-1})}{(1-\alpha)(1+\alpha)^{r-1}}F\nonumber \\
 & =\frac{(1+\alpha+\ldots\alpha^{r-1})}{(1+\alpha)^{r-1}}F\nonumber \\
 & \leq F.
\end{align}

Therefore, we can find that the reward can not exceed the transaction fee paid by the \emph{Tx\_msg }producer. Proposition 1 is proven.
\end{IEEEproof}

\begin{algorithm}[t]
\textbf{Input:} The relay action set $\boldsymbol{a}$, and weight vector $\mathbf{w}$.

\textbf{Output:} The relaying probability vector $\mathbf{p}$.

\ \ 1: \textbf{Initialization}

\ \ 2: \ \ \ \  $w_{a}^{1}\leftarrow1$

\ \ 3: \ \ \ \  $\varGamma{}^{1}\leftarrow\sum_{a}w_{a}^{1}$

\ \ 4: \ \ \ \  $p^{1}\left(a\right)\leftarrow\nicefrac{w_{a}^{1}}{\varGamma^{1}}$

\ \ 5: \textbf{begin}

\ \ 6: \textbf{\ \ \ \  for} each time step t \textbf{do}

\ \ 7: \ \ \ \  \ \ \ \  Caculate the action cost $c^{t}(a)$.

\ \ 8: \ \ \ \  \ \ \ \  $w_{a}^{t+1}\leftarrow w_{a}^{t}\left(1-\gamma c^{t}(a)\right)$

\ \ 9: \ \ \ \  \ \ \ \  $\varGamma{}^{t+1}\leftarrow\sum_{a}w_{a}^{t+1}$

\hspace{0.05em}10: \ \ \ \  \ \ \ \  $p^{t+1}\left(a\right)\leftarrow\nicefrac{w_{a}^{t+1}}{\varGamma^{t+1}}$

\hspace{0.05em}11: \ \ \textbf{\ \  end for}

\hspace{0.05em}12: \textbf{end}

\caption{No-Regret Algorithm}
\end{algorithm}

\subsection{First Stage: Relay Sub-auction}

As shown in Fig. 4, when relay nodes receive \emph{Block\_msg}, they append it to the blockchain and adjust the relaying probability. While receiving the \emph{Tx\_msg}, relay nodes first judge whether put it into the transaction pool. Then, nodes choose \emph{Tx\_msg}s through the relay sub-auction. At last, relay nodes use goosip protocol to propagate the winning \emph{Tx\_msg}s. The relay sub-auction is a double auction. Transaction producers attach transaction fees to bid, and relay nodes sell their energy and bandwidth to route \emph{Tx\_msg}s.

\begin{figure}[t]
\includegraphics{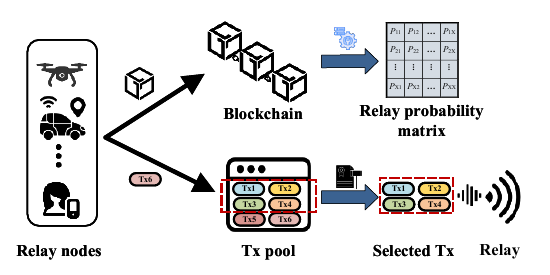}
\caption{The process of relay sub-auction.}
\end{figure}

\begin{algorithm}[th]
\textbf{Input:} Received messages $m$, relay node's transaction pool $\mathcal{P}$, blockchain $\mathcal{B}$ and cost $c_{f}$.

\textbf{Output:} Relay message set $\mathcal{M}$.

\ \ 1: \textbf{begin}

\ \ 2: \ \ \ \  \textbf{if} $m$ is the \emph{Block\_msg} \textbf{then}

\ \ 3: \ \ \ \  \ \ \ \  \textbf{if} \emph{Block\_msg }is valid \textbf{then}

\ \ 4: \ \ \ \  \ \ \ \  \ \ \ \  $\mathcal{B\leftarrow B\ \cup}$ \emph{Block\_msg}

\ \ 5: \ \ \ \  \ \ \ \  \ \ \ \  Broadcast \emph{Block\_msg}

\ \ 6: \ \ \ \  \ \ \ \  \ \ \ \  \emph{$\boldsymbol{\triangleright}$}\textbf{No-Regret Algorithm}

\ \ 7: \ \ \ \  \ \ \ \  \textbf{end if}

\ \ 8: \ \ \ \  \textbf{else if} $m$ is the \emph{Tx\_msg} \textbf{then}

\ \ 9: \ \ \ \  \ \ \ \  \textbf{if} $l_{t}\geq1$ \textbf{and} $rnd\leq p$ \textbf{then}

\hspace{0.05em}10: \ \ \ \  \ \ \ \  \ \ \ \  $w_{t}\leftarrow\min\left(1,\frac{T_{c}}{T-T_{s}}\right)$

\hspace{0.05em}11: \ \ \ \  \ \ \ \  \ \ \ \  $\widetilde{\nu}\leftarrow w_{t}(\frac{1}{1+\alpha})^{L+1-l_{t}}\alpha^{L+1-l_{t}}F$

\hspace{0.05em}12: \ \ \ \  \ \ \ \  \ \ \ \  \textbf{if} $\widetilde{\nu}-c_{f}\geq0$ \textbf{then}

\hspace{0.05em}13: \ \ \ \  \ \ \ \  \ \ \ \  \ \ \ \ $\mathcal{X}\leftarrow\varnothing$

14: \ \ \ \  \ \ \ \  \ \ \ \  \ \ \ \  $\mathcal{X}\leftarrow$\emph{Tx\_msg}

\hspace{0.05em}15: \ \ \ \  \ \ \ \  \ \ \ \  \ \ \ \ \textbf{for} $i$ \textbf{in} $\mathcal{P}$ \textbf{do}

\hspace{0.05em}16: \ \ \ \  \ \ \ \  \ \ \ \  \ \ \ \ \ \ \ \  \textbf{if} $i\oplus$\emph{Tx\_msg $\equiv1$ }\textbf{then}

\hspace{0.05em}17: \ \ \ \  \ \ \ \  \ \ \ \  \ \ \ \ \ \ \ \  \ \ \ \  $\mathcal{X}\leftarrow i,\mathcal{P}\leftarrow\mathcal{P}\backslash\{i\}$

\hspace{0.05em}18: \ \ \ \  \ \ \ \  \ \ \ \  \ \ \ \ \ \ \ \  \textbf{end if}

\hspace{0.05em}19: \ \ \ \  \ \ \ \  \ \ \ \  \ \ \ \ 
\textbf{end for}

\hspace{0.05em}20: \ \ \ \  \ \ \ \  \ \ \ \  \ \ \ \ $j\leftarrow\arg\max_{j\in\mathcal{X}}\widetilde{\nu}_{j}$

\hspace{0.05em}21: \ \ \ \  \ \ \ \  \ \ \ \  \ \ \ \ $\mathcal{P}\leftarrow\mathcal{P}\cup\{j\}$

\hspace{0.05em}22: \ \ \ \  \ \ \ \  \ \ \ \  \textbf{end if}

\hspace{0.05em}23: \ \ \ \  \ \ \ \  \textbf{end if}

\hspace{0.05em}24: \ \ \ \  \textbf{end if}

\hspace{0.05em}25: \ \ \ \  \textbf{while} $\mid\mathcal{P}\mid$$>N$ \textbf{do}

\hspace{0.05em}26: \ \ \ \  \ \ $\mathbf{\#}$ Determine winner

\hspace{0.05em}27: \ \ \ \  \ \ \ \  Sort $\mathcal{P}$ in descending order $\left\{ \widetilde{\nu}_{1},\ldots,\widetilde{\nu}_{N},\ldots,\widetilde{\nu}_{\mid\mathcal{P}\mid}\right\} $

\hspace{0.05em}28: \ \ \ \  \ \ \ \  $\mathcal{M}\leftarrow\varnothing $

\hspace{0.05em}29: \ \ \ \  \ \ \ \  \textbf{for} $k$ \textbf{in} $\mathcal{P}$ \textbf{do}

\hspace{0.05em}30: \ \ \ \  \ \ \ \  \ \ \ \  $\mathcal{M}\leftarrow k,\mathcal{P\leftarrow P\backslash}\{k\}$

\hspace{0.05em}31: \ \ \ \  \ \ \ \  \ \ \ \  \textbf{if} $\mid\mathcal{M}\mid\geq N$ \textbf{then}

\hspace{0.05em}32: \ \ \ \  \ \ \ \  \ \ \ \  \ \ \ \ \textbf{break}

\hspace{0.05em}33: \ \ \ \  \ \ \ \  \ \ \ \  \textbf{end if}

\hspace{0.05em}34: \ \ \ \  \ \ \ \  \textbf{end for}

\hspace{0.05em}35: \ \ \ \  \textbf{end while}

\hspace{0.05em}36: \ \ \ \  \textbf{for} $T_{x}$ \textbf{in} $\mathcal{M}$ \textbf{do}

\hspace{0.05em}37: \ \ \ \  \ \ \ \  Sign $\left\langle I,g\right\rangle$ on $T_{x}$

\hspace{0.05em}38: \ \ \ \  \ \ \ \  $l_{t}\leftarrow l_{t}-1,$ broadcast $T_{x}$

\hspace{0.05em}39: \ \ \ \  \textbf{end for}

\hspace{0.05em}40: \textbf{end}

\caption{Relay Sub-auction}
\end{algorithm}

When the \emph{Tx\_msg }producer broadcasts the message, it attaches the bid $\left\langle F,T_{c},T_{s},l_{t}\right\rangle $, where $F$ is the cap of the paid transaction fee, $T_{c}$ is the expected confirmation time of the \emph{Tx\_msg}, $T_{s}$ is the timestamp and $l_{t}$ is the remaining times of relaying. Other relay nodes relay the message according to the relaying probability $p$ while receiving the \emph{Tx\_msg}. The relaying probability $p$ is vital for the wireless blockchain system to reduce energy and resource consumption and confirmed delay. Considering the distributed environment of blockchain, we use the no-regret algorithm \cite{roughgarden2016twenty}, \cite{cesa2006prediction} to adjust the relaying probability $p$. 

The no-regret algorithm is shown in Algorithm 1. While receiving the \emph{Tx\_msg}, the relay node has two actions to choose from, $a_{0}$ means discard the message, and $a_{1}$ means relay it. $w_{a}$ is the weight of the action, and $p(a)$ is the relaying probability of the action $a$. During the initialization, both probabilities of actions are set to $\frac{1}{2}$ (lines 1-4). Even though the relay node relays the \emph{Tx\_msg}, it may not obtain the reward. On the one hand, the \emph{Tx\_msg} may not reach full nodes due to the complex wireless environment. On the other hand, full nodes select one \emph{Tx\_msg} for the same transaction, and other \emph{Tx\_msg}s relayed by different routings are discarded. Therefore, the relay node needs to adjust the relaying probability based on confirmation results of \emph{Tx\_msg}s on the blockchain. Each receives a valid block, and the relay node adjusts the probability according to the block's transaction (lines7-10), where $\gamma$ is the learning rate to adjust exploration and development, and $\varGamma$ is the sum of action weights. $c(a)$ is the cost of the action, and we define it as the negative utility. If the relay node chooses to relay the message, there are two results: \emph{Tx\_msg} is confirmed successfully with $c(a_{1})=c_{f}-\nu$, or failed with $c(a_{1})=c_{f}$, where $\nu$ is paid reward. Suppose the light node discards the message, then $c(a_{0})=0$. By iterating, the no-regret algorithm could converge the probability distribution to the Coarse Correlated Equilibria (CCE) \cite{hannan1957approximation}.

Algorithm 2 demonstrates the process of the relay sub-auction. If the relay node receives \emph{Block\_msg}, it verifies the message's validity and then adjusts the relaying probability set according to Algorithm 1 (lines 2-6). If the message is \emph{Tx\_msg}, the relay node first decides whether to receive it based on $l_{t}$ and the relaying probability $p$, where $rnd$ is a random number between 0 and 1. The transaction waiting time is a key factor for relay nodes, transaction producers are willing to pay more fees to shorten their transaction waiting time. To accelerate \emph{Tx\_msg} relay, we introduce the time weight $w_{t}$ to relate the reward to waiting time, where $T$ is the current time, $T_{s}$ is the timestamp that produced the \emph{Tx\_msg}, and $T_{c}$ is the expected delay of confirmation (lines 10-11). The confirmation delay is shorter, the relay nodes will receive more transaction fees. Then, the relay node estimates the reward using the path length $L+2-l_{t}$ since the node is unaware of the final path length, where $L$ is the maximum length of the relay path. We define the operator $\oplus$ to judge whether the two messages are for the same transaction. To maximize utility, the relay node only forwards the \emph{Tx\_msg} of the maximum reward for the same transaction (lines 15-21). Once the number of \emph{Tx\_msg}s in the transaction pool $\mathcal{P}$ exceeds $N$, the relay node selects the first $N$ highest reward \emph{Tx\_msg}s to broadcast (lines 25-38). The relay node signs $\left\langle I,g\right\rangle$ on the \emph{Tx\_msg}, where $I$ is the identity and $g$ is the relay order. Besides, the relay node uses a digital signature to prevent the relay list from being changed.

\subsection{Second Stage: Validation Sub-auction}

Fig. 5 depicts the process of the validation sub-auction. According to the VCG mechanism, full nodes select victorious \emph{Tx\_msg}s and calculate the reward. Then, full nodes package the \emph{Tx\_msg}s to the block and attach the validation reward. While resolving the hash puzzle, full nodes broadcast the new block to the wireless blockchain network. The validation sub-aution is also a double auction. Transaction producers pay fees for \emph{Tx\_msg} confirmation, and full nodes consume computing resource to validate and package \emph{Tx\_msg}s.

\begin{figure}[t]
\includegraphics[scale=0.82]{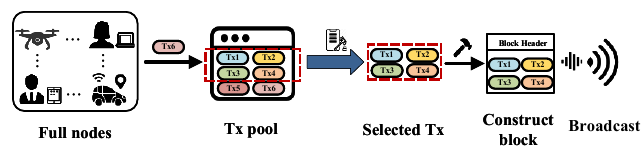}
\caption{The process of validation sub-auction.}
\end{figure}

As illustrated in Algorithm 3, the full node first judges the message type. Since the full node is the destination of the \emph{Tx\_msg}, it can obtain the exact path length to calculate the reward. Similar to relay sub-auction, the full node also selects the \emph{Tx\_msg} of the maximum reward for the same transaction (lines 13-19). We apply the VCG mechanism \cite{krishna2009auction} to determine the winning \emph{Tx\_msg}s and the paid reward $\nu$ (lines 32). The \emph{Block\_msg} consists of $\left\langle H,I,Nonce,\mathcal{L},T_{B},\nu\right\rangle$, where $H$ is the block height, $I$ is the identity, $Nonce$ is a 4-Byte number, $\mathcal{L}$ is the list of \emph{Tx\_msg}s, $T_{B}$ is the block generation time, and $\nu$ is the paid reward for the full node. The full node keeps trying $Nonce$ to satisfy the hash target while building the \emph{Block\_msg} (lines 32-38).

\begin{algorithm}[t]
\textbf{Input:} Received messages $m$, full node's transaction pool $\mathcal{P}$, blockchain $\mathcal{B}$ and cost $c_{v}$. 

\textbf{Output:} \emph{Block\_msg} $\left\langle H,I,Nonce,\mathcal{L},T_{B},\nu\right\rangle $.

\ \ 1: \textbf{begin}

\ \ 2: \ \ \ \  \textbf{if} $m$ is the \emph{Block\_msg} \textbf{then}

\ \ 3: \ \ \ \  \ \ \ \  \textbf{if} \emph{Block\_msg }is valid \textbf{then}

\ \ 4: \ \ \ \  \ \ \ \  \ \ \ \  $\mathcal{B\leftarrow B\ \cup}$ \emph{Block\_msg}

\ \ 5: \ \ \ \  \ \ \ \  \ \ \ \  Broadcast \emph{Block\_msg}

\ \ 6: \ \ \ \  \ \ \ \  \textbf{end if}

\ \ 7: \ \ \ \  \textbf{else if} $m$ is the \emph{Tx\_msg} \textbf{then}

\ \ 8: \ \ \ \  \ \ \ \  $w_{t}\leftarrow\min\left(1,\frac{T_{c}}{T-T_{s}}\right)$

\ \ 9: \ \ \ \  \ \ \ \ \  $\widetilde{\nu}\leftarrow w_{t}(\frac{1}{1+\alpha})^{L-l_{t}}F$

\hspace{0.05em}10:\textbf{ \ \ \ \  \ \ \ \  if} $\widetilde{\nu}-c_{v}\geq0$ \textbf{then}

\hspace{0.05em}11: \ \ \ \  \ \ \ \  \ \ \ \  $\mathcal{X}\leftarrow\varnothing$

12: \ \ \ \  \ \ \ \  \ \ \ \  $\mathcal{X}\leftarrow$\emph{Tx\_msg}

\hspace{0.05em}13: \ \ \ \  \ \ \ \  \ \ \ \  \textbf{for} $i$ \textbf{in} $\mathcal{P}$ \textbf{do}

\hspace{0.05em}14: \ \ \ \  \ \ \ \  \ \ \ \  \ \ \ \ \textbf{if} $i\oplus$\emph{Tx\_msg $\equiv1$ }\textbf{then}

\hspace{0.05em}15: \ \ \ \  \ \ \ \  \ \ \ \  \ \ \ \ \ \ \ \  $\mathcal{X}\leftarrow i,\mathcal{P}\leftarrow\mathcal{P}\backslash\{i\}$

\hspace{0.05em}16: \ \ \ \  \ \ \ \  \ \ \ \  \ \ \ \ \textbf{end if }

\hspace{0.05em}17: \ \ \ \  \ \ \ \  \ \ \ \  \textbf{end for}

\hspace{0.05em}18: \ \ \ \  \ \ \ \  \ \ \ \  $j\leftarrow\arg\max_{j\in\mathcal{X}}\widetilde{\nu}_{j}$

\hspace{0.05em}19: \ \ \ \  \ \ \ \  \ \ \ \  $\mathcal{P}\leftarrow\mathcal{P}\cup\{j\}$

\hspace{0.05em}20: \ \ \ \  \ \ \ \  \textbf{end if}

\hspace{0.05em}21: \ \ \ \  \textbf{end if}

\hspace{0.05em}22: \ \ \ \  \textbf{while} $\mid\mathcal{P}\mid$$>M$ \textbf{do}

\hspace{0.05em}23: \ \ \ \  \ \ $\mathbf{\#}$VCG mechanism

\hspace{0.05em}24: \ \ \ \  \ \ \ \  Sort $\mathcal{P}$ in descending order $\left\{ \widetilde{\nu}_{1},\ldots,\widetilde{\nu}_{M+1},\ldots,\widetilde{\nu}_{\mid\mathcal{P}\mid}\right\}$

\hspace{0.05em}25: \ \ \ \  \ \ \ \  $\mathcal{L}\leftarrow\varnothing $

\hspace{0.05em}26: \ \ \ \  \ \ \ \  \textbf{for} $k$ \textbf{in} $\mathcal{P}$ \textbf{do}

\hspace{0.05em}27: \ \ \ \  \ \ \ \  \ \ \ \  $\mathcal{L}\leftarrow k,\mathcal{P\leftarrow P\backslash}\{k\}$

\hspace{0.05em}28: \ \ \ \  \ \ \ \  \ \ \ \  \textbf{if} $\mid\mathcal{L}\mid\geq M$ \textbf{then}

\hspace{0.05em}29: \ \ \ \  \ \ \ \  \ \ \ \  \ \ \ \ \textbf{break}

\hspace{0.05em}30: \ \ \ \  \ \ \ \  \ \ \ \  \textbf{end if}

\hspace{0.05em}31: \ \ \ \  \ \ \ \  \textbf{end for}

\hspace{0.05em}32: \ \ \ \  \textbf{end while }

\hspace{0.05em}33: \ \ \ \  $\nu\leftarrow$$\widetilde{\nu}_{M+1}$, $T_{B}\leftarrow T_{c}$

\hspace{0.05em}34: \ \ \ \  Produce \emph{Block\_msg} $\left\langle H,I,Nonce,\mathcal{L},T_{B},\nu\right\rangle $

\hspace{0.05em}35: \ \ \ \  \textbf{while} $Hash$(\emph{Block\_msg}) $>$ \emph{traget} \textbf{do}

\hspace{0.05em}36: \ \ \ \  \ \ \ \  $Nonce\leftarrow Nonce^{'}$, $T_{B}\leftarrow T$

\hspace{0.05em}37: \ \ \ \  \ \ \ \  Rebuild \emph{Block\_msg} $\left\langle H,I,Nonce,\mathcal{L},T_{B},\nu\right\rangle$

\hspace{0.05em}38: \ \ \ \  \textbf{end while}

\hspace{0.05em}39: \ \ \ \  Broadcast \emph{Block\_msg} $\left\langle H,I,Nonce,\mathcal{L},T_{B},\nu\right\rangle $

\hspace{0.05em}40: \textbf{end}

\caption{Validation Sub-auction}
\end{algorithm}

In Algorithm 3, we give the whole designed dual auction mechanism. Each time a valid block is proposed, the dual auction mechanism determines the winners and payments (lines 7-18). Through the designed mechanism, light and full nodes are willing to receive the \emph{Tx\_msg} with the shortest path and minimum delay. Thus, our mechanism reduces resource consumption and helps find the optimal routing between \emph{Tx\_msg} producers and full nodes.

\begin{algorithm}[t]
\textbf{Input:} Messages $\mathbf{m}$, light nodes $\mathbf{x}$ and full nodes $\mathbf{y}$.

\textbf{Output:} The winner $\boldsymbol{\mathbf{\mathbf{\boldsymbol{\omega}}}}$ and payment $\boldsymbol{\boldsymbol{\pi}}$.

\ \ 1: \textbf{begin}

\ \ 2: \ \ \ \  \textbf{if} node is the light node \textbf{then}

\ \ 3: \ \ \ \  \ \ \ \  \textbf{$\boldsymbol{\triangleright}$ Relaying Sub-auction}

\ \ 4: \ \ \ \  \textbf{else if} node is the full node \textbf{then}

\ \ 5: \ \ \ \  \ \ \ \  \textbf{$\boldsymbol{\triangleright}$ Relaying Sub-auction}

\ \ 6: \ \ \ \  \ \ \ \  \textbf{$\boldsymbol{\triangleright}$ Validation Sub-auction}

\ \ 7: \ \ \ \  \textbf{end if}

\ \ 8: \ \ \ \  \textbf{for }each valid \emph{Block\_msg} \textbf{do}

\ \ 9: \ \ \ \  \ \ \ \  $\pi_{v}\leftarrow\nu$

\hspace{0.05em}10: \ \ \ \  \ \ \ \  $\omega_{v}\leftarrow I_{v}$

\hspace{0.05em}11: \ \ \ \  \ \ \ \  \textbf{for $i$ in} $\mathcal{L}$ \textbf{do}

\hspace{0.05em}12:\textbf{ \ \ \ \  \ \ \ \  \ \ \ \ $w_{t}^{i}\leftarrow\min\left(1,\frac{T_{c}}{T_{B}-T_{s}}\right)$}

\hspace{0.05em}13:\textbf{ \ \ \ \  \ \ \ \  \ \ \ \ $\rho_{i}\leftarrow w_{t}^{i}(\frac{1}{1+\alpha})^{r_{i}-1}F_{i}$}

\hspace{0.05em}14:\textbf{ \ \ \ \  \ \ \ \  \ \ \ \ $\boldsymbol{\pi}_{i}\leftarrow\left\{ \alpha^{1}\rho_{i},\ldots,\alpha^{r_{i}-1}\rho_{i}\right\}$}

\hspace{0.05em}15:\textbf{ \ \ \ \  \ \ \ \  \ \ \ \ $\boldsymbol{\omega}_{i}\leftarrow\left\{ I_{1},\ldots,I_{r_{i}-1}\right\}$}

\hspace{0.05em}16:\textbf{ \ \ \ \  \ \ \ \  end for}

\hspace{0.05em}17: \ \ \ \  \ \ \ \  $\boldsymbol{\boldsymbol{\pi}}\leftarrow\left\{ \pi_{v},\boldsymbol{\pi}_{1},\ldots,\boldsymbol{\pi}_{\mid\mathcal{L}\mid}\right\}$

\hspace{0.05em}18: \ \ \ \  \ \ \ \  $\boldsymbol{\boldsymbol{\omega}}\leftarrow\left\{ \omega_{v},\boldsymbol{\omega}_{1},\ldots,\boldsymbol{\omega}_{\mid\mathcal{L}\mid}\right\}$

\hspace{0.05em}19:\textbf{ \ \ \ \  end for}

\hspace{0.05em}20:\textbf{ end}

\caption{Dual Auction Mechanism}
\end{algorithm}

According to the dual auction mechanism, we rewrite the utility function of light and full nodes as follows:

\begin{align}
U_{f}^{i} & =\sum_{j}\mathop{\left(\eta_{ij}w_{t}^{ij}(\frac{1}{1+\alpha})^{r_{j}-1}\alpha^{g_{ij}}F_{j}-p_{ij}c_{f}^{i}\right)}\nonumber \\
 & \ \ \ \ +\sum_{k}\delta_{k}^{i}\left(\mathop{V_{k}^{i}-F_{k}^{i}}\right),
\end{align}

\begin{align}
U_{v}^{l} & =\sum_{j}\mathop{\left(\eta_{lj}w_{t}^{lj}(\frac{1}{1+\alpha})^{r_{j}-1}\alpha^{g_{lj}}F_{j}-p_{lj}c_{f}^{l}\right)}\nonumber \\
 & \ \ \ \ +\sum_{s}\left(\sigma_{s}^{l}\sum_{n}w_{t}^{s}(\frac{1}{1+\alpha})^{r_{s}-1}F_{s}-c_{v}^{l}\right).
\end{align}

To maximize social welfare, we should determine the relaying probability $p$ and transaction fee $F$. Through the dual auction mechanism, the relaying probability $p$ converges to the CCE. In the next section, we prove that the mechanism guarantees that the honest bid $F=V$ is optimal.

\section{Performance Analysis}

This section analyzes the IR, IC, CE, and upper regret bound of the dual auction mechanism. 

\subsection{Individual Rationality}
\begin{thm}
The dual auction mechanism is individually rational for both light and full nodes.
\end{thm}
\begin{IEEEproof}
From Algorithm 3, the full node selects the transaction with the reward that exceeds the computing cost. Besides, a \emph{Block\_msg} includes $M$ \emph{Tx\_msg}s to ensure the full node obtains multiple rewards. Once a full node proposes a block successfully, it will receive positive rewards. From the perspective of light nodes, even though they choose to relay the reward of \emph{Tx\_msg} exceeding the cost, they may gain a negative reward due to the failure of \emph{Tx\_msg} confirmation.
Therefore, we prove the light node will receive a non-negative reward through the dual auction mechanism.

Section 4 dedicates that the relaying probability $p$ converges the CCE. Based on the definition of CCE, we derive

\begin{equation}
\boldsymbol{E}_{\boldsymbol{a\sim p}}\left[u_{i}(\boldsymbol{a})\right]\geq\boldsymbol{E}_{\boldsymbol{a\sim p}}\left[u_{i}(a_{i}^{'},\boldsymbol{a_{-i}})\right],
\end{equation}
where $\boldsymbol{a}$ is the action set, $u_{i}(\boldsymbol{a})$ is the utility of relay node $i$ based on the relaying probability distribution $\boldsymbol{p}$, and $\boldsymbol{E}_{\boldsymbol{a\sim p}}\left[u_{i}(\boldsymbol{a})\right]$ is the expected utility. If the node does not relay the message, it has no cost or reward. We fix $a_{i}^{'}=a_{0}$ and obtain the following equation

\begin{equation}
\boldsymbol{E}_{\boldsymbol{a\sim p}}\left[u_{i}(\boldsymbol{a})\right]\geq\boldsymbol{E}_{\boldsymbol{a\sim p}}\left[u_{i}(a_{0},\boldsymbol{a_{-i}})\right]=0.
\end{equation}

Thus, the light node can obtain the non-negative expected utility by dual auction, and Theorem 2 is proved.
\end{IEEEproof}

\subsection{Incentive Compatibility}
\begin{thm}
The dual auction mechanism is incentive compatible.
\end{thm}
\begin{IEEEproof}
To guarantee the IC of the dual auction, we should prove that the light node will honestly bid while producing the \emph{Tx\_msg}. For convenience, we call the \emph{Tx\_msg} producer the bidder and the transaction fee as the bid. From Section 4, we derive that the designed mechanism comprises the relay sub-auction and validation sub-auction. Nevertheless, only the validation sub-auction can determine winners among bidders. If the validation sub-auction is Dominant-Strategy Incentive Compatible (DSIC), then the dual auction mechanism is DSIC. We define the utility of the bidder $i$ in the validation sub-auction as follows

\begin{equation}
u_{i}=V_{i}(\boldsymbol{\omega})-P_{i}(\boldsymbol{\omega}),\label{eq:utility of mining auction}
\end{equation}
where $\boldsymbol{\mathbf{\omega}}$ is the winner set, $V_{i}$ is the value of the transaction, and $P_{i}$ is the payment of the validation sub-auction.

Based on Algorithm 4, the allocation rule is given

\begin{equation}
\boldsymbol{\omega^{*}}=\arg\max_{\boldsymbol{\boldsymbol{\omega}}\in\Omega}\sum_{i=1}^{M}b_{i}(\boldsymbol{\omega}),
\end{equation}
where $b$ is the bid, and $\Omega$ is all possible allocation schemes. It is easy to deduce that the allocation rule $\boldsymbol{\omega}^{\boldsymbol{*}}$ is monotonous.

According to the VCG mechanism, the payment rule is 

\begin{equation}
P_{i}(\boldsymbol{\omega^{*}})=b_{i}(\boldsymbol{\omega^{*}})-\left[\sum_{j=1}^{M}b_{j}(\boldsymbol{\omega^{*}})-\max_{\boldsymbol{\boldsymbol{\omega}}}\sum_{j\neq i}b_{j}(\boldsymbol{\omega})\right].\label{eq:payment rule}
\end{equation}

Substitute equation \eqref{eq:payment rule} into equation \eqref{eq:utility of mining auction}, we rewrite the utility function 

\begin{equation}
u_{i}=V_{i}(\boldsymbol{\omega^{*}})+\sum_{j\neq i}b_{j}(\boldsymbol{\boldsymbol{\omega^{*}}})-\max_{\boldsymbol{\boldsymbol{\omega}}}\sum_{j\neq i}b_{j}(\boldsymbol{\omega}).
\end{equation}

Next, we prove that the bidder who bids honestly will gain the maximum utility. From Algorithm 3, a crucial bid $b_{c}$ (i.e., $\widetilde{\nu}_{M+1}$) exists, and other bids above it win the auction. 

1) $V_{i}>b_{c}$: If the bidder $i$ bids honestly $b_{i}=V_{i}$, it will win the auction and pay $P_{i}=b_{c}$. The utility of bidder $i$ is $V_{i}-b_{c}\geq0$. In the case of $b_{i}\neq V_{i}$, if $b_{i}>b_{c}$, the bidder wins the auction and obtains the utility $V_{i}-b_{c}$. However, if $b_{i}\leq b_{c}$, bidder $i$ loses the auction and receives zero utility. The honest bid is the dominant strategy.

2) $V_{i}\leq b_{c}$: If the bidder $i$ bids $b_{i}=V_{i}$ honestly, it will lose the auction and obtain zero utility. In the case of submitting a dishonest bid $b_{i}\neq V_{i}$, the bidder gains zero utility while $b_{i}\leq b_{c}$. Or the bidder $i$ submit bid $b_{i}>b_{c}$, it wins the auction and obtains negative utility $V_{i}-b_{c}<0$. The honest bid also brings maximal utility.

Therefore, the designed auction mechanism is DSIC, and Theorem 3 is proved.
\end{IEEEproof}

\subsection{Computational Efficiency}
\begin{thm}
The dual auction mechanism is computationally efficient.
\end{thm}
\begin{IEEEproof}
In Algorithm 2, to eliminate the redundant \emph{Tx\_msg}s, the process (lines 15-21) has the time complexity of $O(N)$. The process of winner selection (lines 25-34) has the time complexity of $O(N^{2}\log N)$. Ultimately, the digital signature (lines 35-38) needs the time complexity of $O(N)$. Algorithm 2 has the time complexity of $O(N^{2}\log N)$. Similarly, Algorithm 3 needs the time complexity of $O(N^{2}\log N)$ to run in total. Therefore, the whole dual auction mechanism is computationally efficient and bounded by the polynomial time complexity of $O(N^{2}\log N)$.
\end{IEEEproof}

\subsection{Upper Regret Bound}

We define regret as the gap between the optimal allocation policy and the dual auction mechanism.
\begin{thm}
For any $\mu\gamma\leq\frac{1}{2}$, the dual auction mechanism has the upper regret bound of $\frac{ln2}{\gamma}+\mu^{2}\gamma K$ in $K$ rounds.
\end{thm}
\begin{IEEEproof}
According to the above analyses, the problem of social welfare maximization is rewritten as follows:

\begin{align}
\max_{\boldsymbol{p}}\sum_{i=1}^{X+Y} & \sum_{j}\mathop{\left(\eta_{ij}w_{t}^{ij}(\frac{1}{1+\alpha})^{r_{j}-1}\alpha^{g_{ij}}V_{j}-p_{ij}c_{f}^{i}\right)}\nonumber \\
 & + \sum_{n=1}^{X}\sum_{k}\delta_{k}^{n}\Bigl(V_{k}^{n}-\Pi\left(\boldsymbol{\nu}_{k}^{n}\right)\Bigr)+\sum_{l=1}^{Y}\sum_{s}\left(\sigma_{s}^{l}M\nu_{m}^{s}-c_{v}^{l}\right)\\
s.t.\nonumber \\
 & \delta_{k}^{i},\eta_{ij},\sigma_{s}^{l}\in\left\{ 0,1\right\} ,\\
 & 0\leq p_{ij}\leq1,
\end{align}
where $\varPi\left(\boldsymbol{\nu}_{k}^{i}\right)$ is the final paid transaction fee of node $i$. Since the honest bid is the dominant strategy for light nodes, we only determine the relaying probability $p$ to maximize social welfare. 

Based on the property of the VCG mechanism, the validation sub-auction satisfies the maximization of social welfare, so we need to analyze the gap between the relay sub-auction and the optimal relay strategy. We define the utility of the node in relay sub-auction during the $\tau$' th round as follows

\begin{equation}
u_{f}^{\tau}=-\sum_{a}p^{\tau}(a)c^{\tau}(a).
\end{equation}

From Algorithm 1, we deduce

\begin{equation}
u_{f}^{\tau}=-\sum_{a}\frac{w_{a}^{\tau}}{\varGamma^{\tau}}c^{\tau}(a),
\end{equation}

\begin{align}
\varGamma^{\tau+1} & =\sum_{a}w_{a}^{\tau+1}=\sum_{a}w_{a}^{\tau}\left(1-\gamma c^{\tau}(a)\right)\nonumber \\
 & =\varGamma^{\tau}(1+\gamma u_{f}^{\tau}).
\end{align}

For $x\in\mathbb{R},1+x\leq e^{x}$ holds, we can derive the following equation

\begin{align}
\varGamma^{K+1} & \leq\varGamma^{K}e^{\gamma u_{f}^{\tau}}\leq\varGamma^{1}\mathop{\prod}_{\tau=1}^{K}e^{\gamma u_{f}^{\tau}}=2\mathop{\prod}_{\tau=1}^{K}e^{\gamma u_{f}^{\tau}},\label{eq:up1}\\
 & \leq2e^{\gamma\sum_{\tau=1}^{K}u_{f}^{\tau}}.
\end{align}

Then, we define the maximum cumulative utility as the optimal utility 

\begin{equation}
u^{*}=\sum_{\tau=1}^{K}u_{f}^{\tau}(a^{*}),
\end{equation}

\begin{equation}
a^{*}=\arg\max_{a}\sum_{\tau=1}^{K}u_{f}^{\tau}(a).
\end{equation}

Based on the nonnegativity of weight $w$, we can obtain

\begin{align}
\varGamma^{\tau+1} & \geq w_{a^{*}}^{\tau+1}=w_{a^{*}}^{1}\prod_{\tau=1}^{K}\left(1+\gamma u_{f}^{\tau}(a^{*})\right)\nonumber \\
 & \geq\prod_{\tau=1}^{K}\left(1+\gamma u_{f}^{\tau}(a^{*})\right).\label{eq:opt}
\end{align}

For convenience, we define $\lambda=\mu\gamma,$ where $\mu$ is the maximum value of $\mid$$u_{f}(a)$$\mid$. We substitute $\lambda$ into the equation \eqref{eq:opt} and obtain

\begin{equation}
\varGamma^{\tau+1}\geq\prod_{\tau=1}^{K}\left(1+\lambda\hat{u}_{f}^{\tau}(a^{*})\right),
\end{equation}
where $\hat{u}_{f}^{\tau}(a^{*})\in\left[-1,1\right]$. Similarly, we use the exponential function to express $1+\lambda\hat{u}_{f}^{\tau}(a^{*})$ approximatively. Through Taylor expansion, we deduce

\begin{align}
\ln(1+x) & =x-\frac{x^{2}}{2}+\frac{x^{3}}{3}-\cdots\nonumber \\
 & \geq x-x^{2},
\end{align}
where $\mid x\mid\leq\frac{1}{2}$. Thus, for any $\mid x\mid\leq\frac{1}{2}$, we have $1+x\geq e^{x-x^{2}}$ and derive the following equation

\begin{equation}
\varGamma^{\tau+1}\geq\prod_{\tau=1}^{K}e^{\lambda\hat{u}_{f}^{\tau}(a^{*})-\lambda^{2}\hat{u}_{f}^{\tau}(a^{*})^{2}}\geq e^{\gamma u^{*}-\lambda^{2}K}.\label{eq:up2}
\end{equation}

According to equations \eqref{eq:up1} and \eqref{eq:up2}, we deduce

\begin{equation}
2e^{\gamma\sum_{\tau=1}^{K}u_{f}^{\tau}}\geq\varGamma^{\tau+1}\geq e^{\gamma u^{*}-\lambda^{2}K}.
\end{equation}

Take the natural logarithm operation and divide by $\gamma$, we obtain

\begin{equation}
u^{*}-\sum_{\tau=1}^{K}u_{f}^{\tau}\leq\frac{ln2}{\gamma}+\mu^{2}\gamma K.
\end{equation}

Therefore, Theorem 5 is proved.
\end{IEEEproof}

\begin{figure*}[t]
    \subfloat[TPS comparision.]{\includegraphics[scale=0.26]{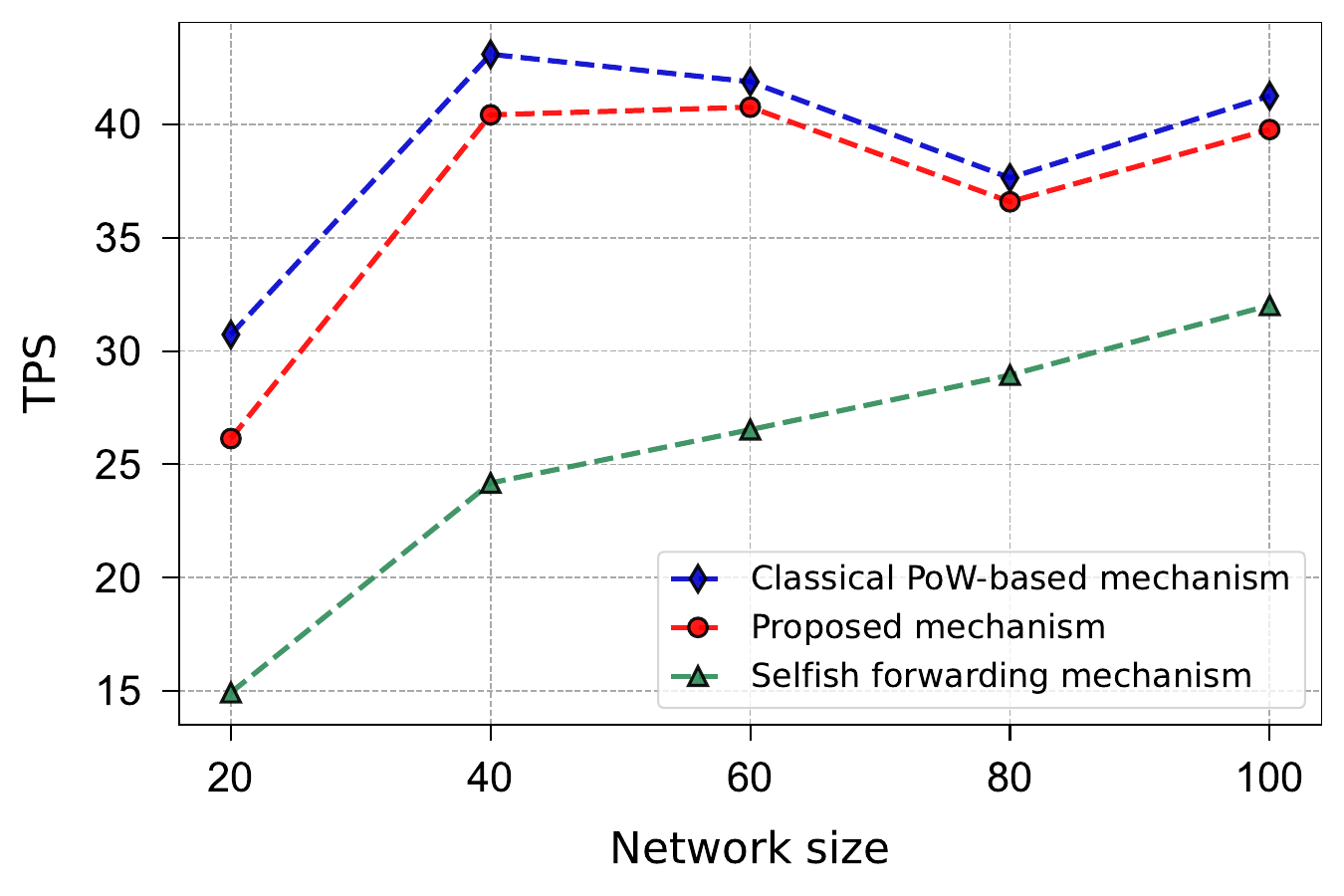}}
    \ \subfloat[Tx confirmation probability.]{\includegraphics[scale=0.26]{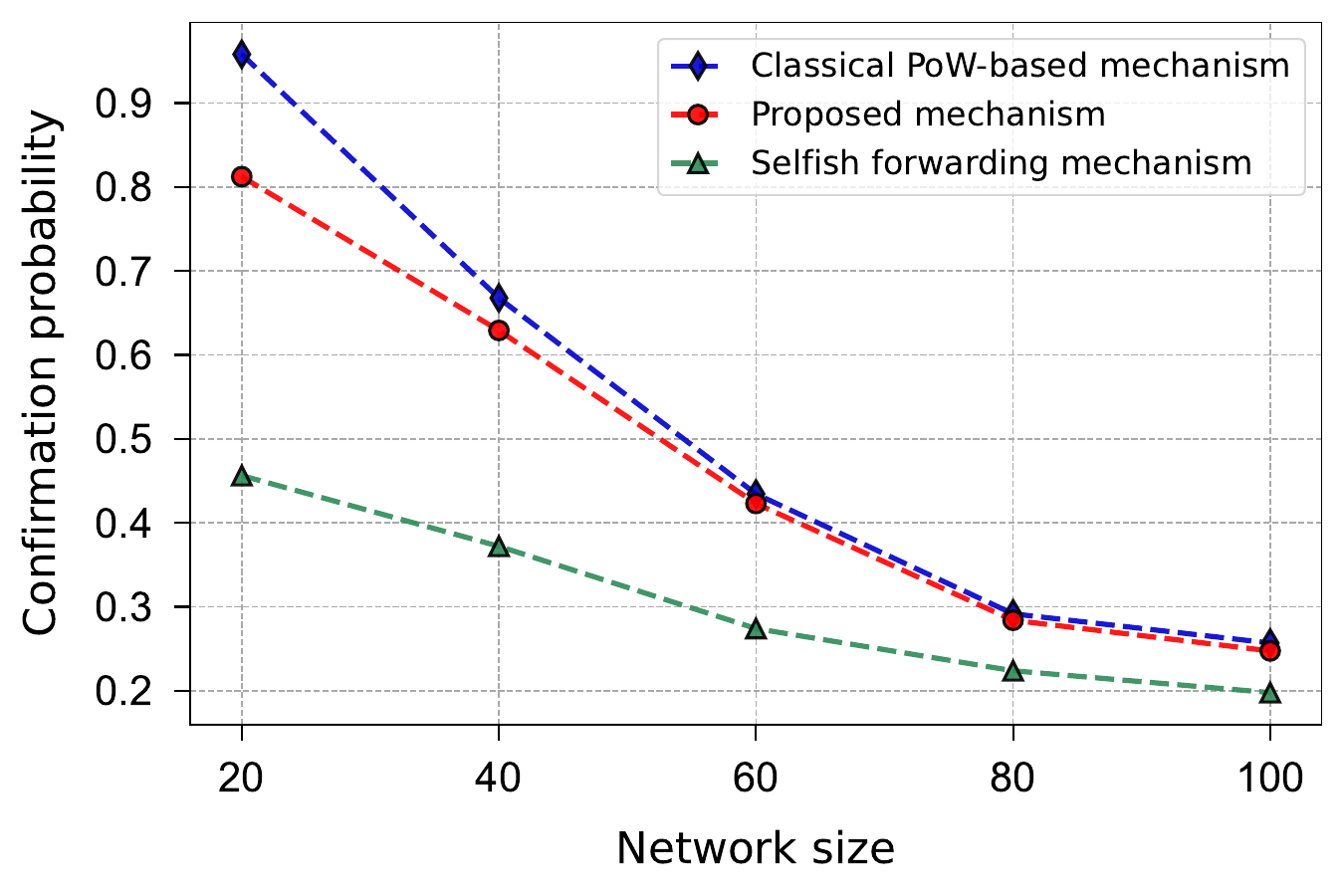}}
    \ \subfloat[Bandwidth consumption.]{\includegraphics[scale=0.26]{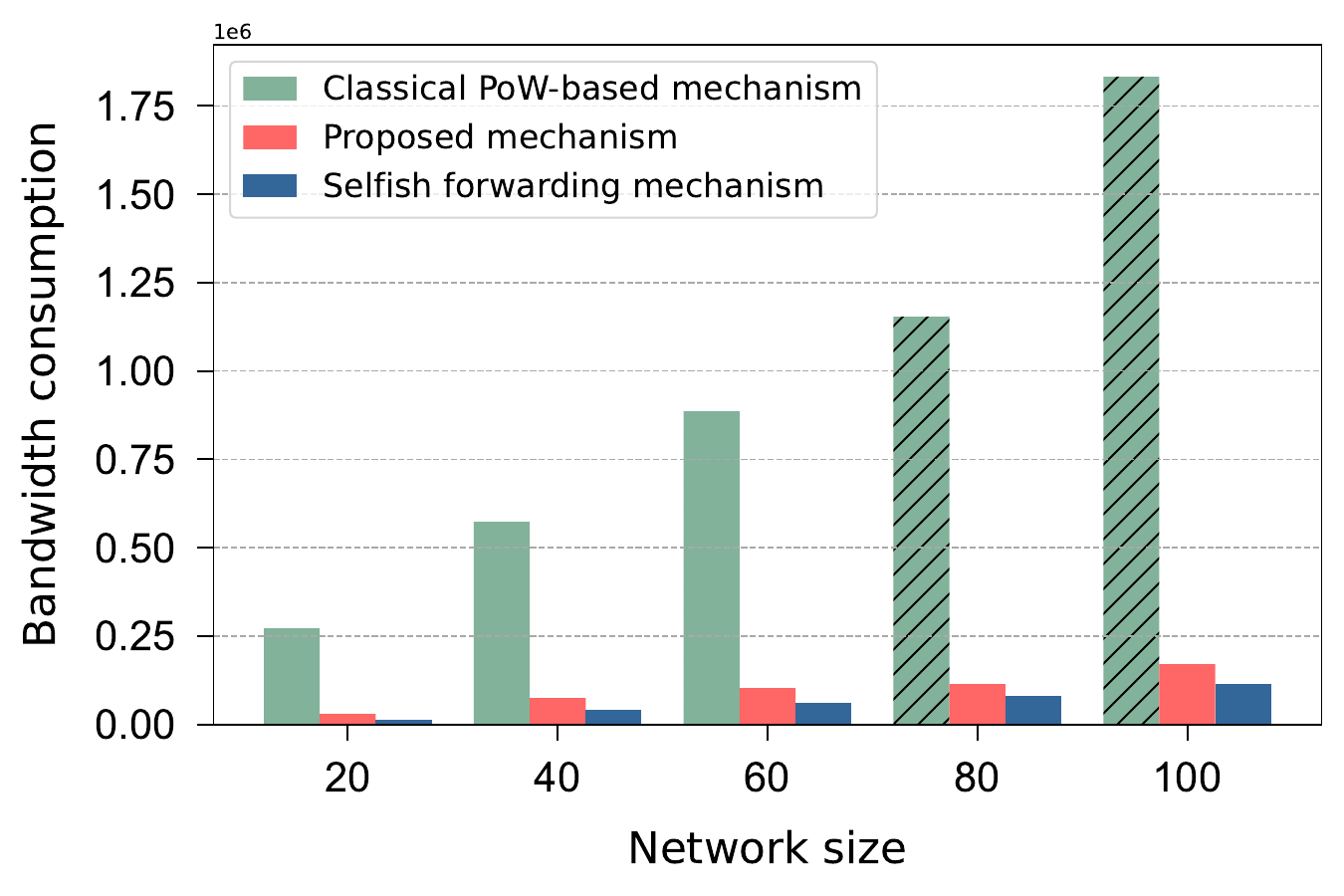}}
    \caption{Running efficiency under different incentive mechanisms in wireless blockchain networks.}
\end{figure*}

\section{Simulation Results }

In this section, we perform simulations to verify the designed dual auction mechanism and compare it with the classical PoW-based and selfish relay mechanisms. The classical PoW-based mechanism represents the existing mechanism in Bitcoin, Ethereum, \emph{etc}. Nodes just transmit their own \emph{Tx\_msgs} under the selfish relay mechanism. Then, we analyze the impact of the dual auction mechanism on social welfare and the performance of the blockchain network. Moreover, we conduct simulations while the wireless blockchain network has faults and malicious nodes and investigate the impact of different complex models on blockchain performance. We set the relay cost, validation cost, and transaction value to follow the normal distribution $\mathcal{N}(1,0.1)$, $\mathcal{N}(5,1)$, and $\mathcal{N}(50,10)$, respectively. Besides, $80$\% of light and $20$\% of full nodes are in the wireless blockchain network. Table 2 lists the default parameters. Not otherwise specified, the parameter is set as the default value.

\begin{table}[tbh]
\caption{Default Parameter Values}
\centering{}%
\begin{tabular}{llll}
\hline 
Parameters & Values\tabularnewline
\hline 
Transaction generation rate & $2$$/s$ \tabularnewline 
Synchronization interval & $10$$s$ \tabularnewline
Block interval & $1$$s$ \tabularnewline
Expected confirmation delay $T_{c}$ & 10$s$ \tabularnewline
Maximum length of relay path $L$ & $6$ \tabularnewline
Number of selected neighbors to forward $n_{f}$ & $3$ \tabularnewline
Maximum number of transactions in a block $M$ & $40$ \tabularnewline
Relay window $N$ & 10 \tabularnewline
Allocation coefficient $\alpha$ & $0.5$ \tabularnewline
Learning rate $\gamma$ & $0.001$ \tabularnewline
\hline 
\end{tabular}
\end{table}

\subsection{Running Efficiency of the Proposed Dual Auction Mechanism}

First, we compare the performance of the proposed dual auction with the Bitcoin and selfish relay mechanisms. Fig. 6(a) illustrates that the selfish relay mechanism has the lowest TPS, and the TPS of the dual auction mechanism is very close to the classical PoW-based mechanism. Since light nodes just relay their transactions under the selfish relay mechanism, full nodes cannot receive enough \emph{Tx\_msg}s to build the block, significantly decreasing the wireless blockchain network's TPS. Nevertheless, there is little TPS decline in the dual auction mechanism compared to the classical PoW-based mechanism. Note that the maximum number of transactions in a block is 40. Thus, when the network size is large, the TPS may be reduced slightly due to
network congestion and limited block size.

In Fig. 6(b), we define the transaction confirmation probability as the ratio of on-chain transactions to total transactions produced by light nodes. Similar to Fig. 6(a), the selfish relay mechanism achieves the worst performance, and the classical PoW-based and dual auction mechanisms get similar confirmation probability. Besides, the most severe problem of the selfish relay mechanism is that only transactions at light nodes connected to full nodes are confirmed, which prevents other light nodes from participating in the wireless blockchain network. Moreover, the transaction confirmation probability of all mechanisms also decreases with the network size increase on account of limited block capacity.

Fig. 6(c) compares the bandwidth consumption among three mechanisms. We assume that relaying a \emph{Tx\_msg} consumes unit bandwidth resources and relaying a \emph{Block\_msg} consumes 40 bandwidth resources. As Fig. 6(c) shows, the classical PoW-based mechanism occupies vast bandwidth resources due to the flooding message relay. The selfish relay mechanism has the minimum information interaction and consumes the fewest bandwidth resources. The dual auction mechanism has bandwidth consumption about twice as the selfish relay mechanism.

\begin{figure*}[t]
    \subfloat[Classical PoW-based mechanism.]{\includegraphics[scale=0.26]{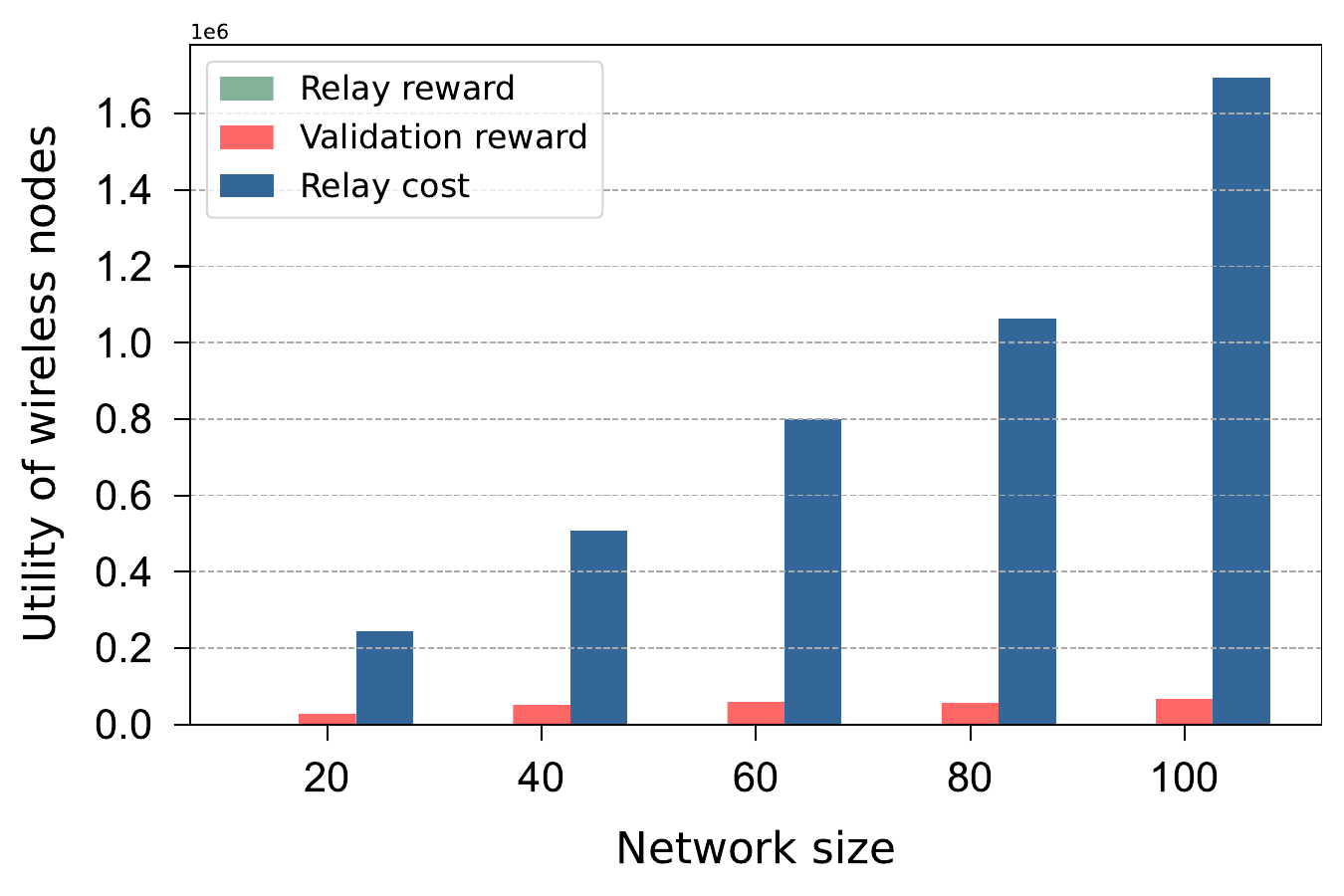}}
    \ \subfloat[Dual auction mechanism.]{\includegraphics[scale=0.26]{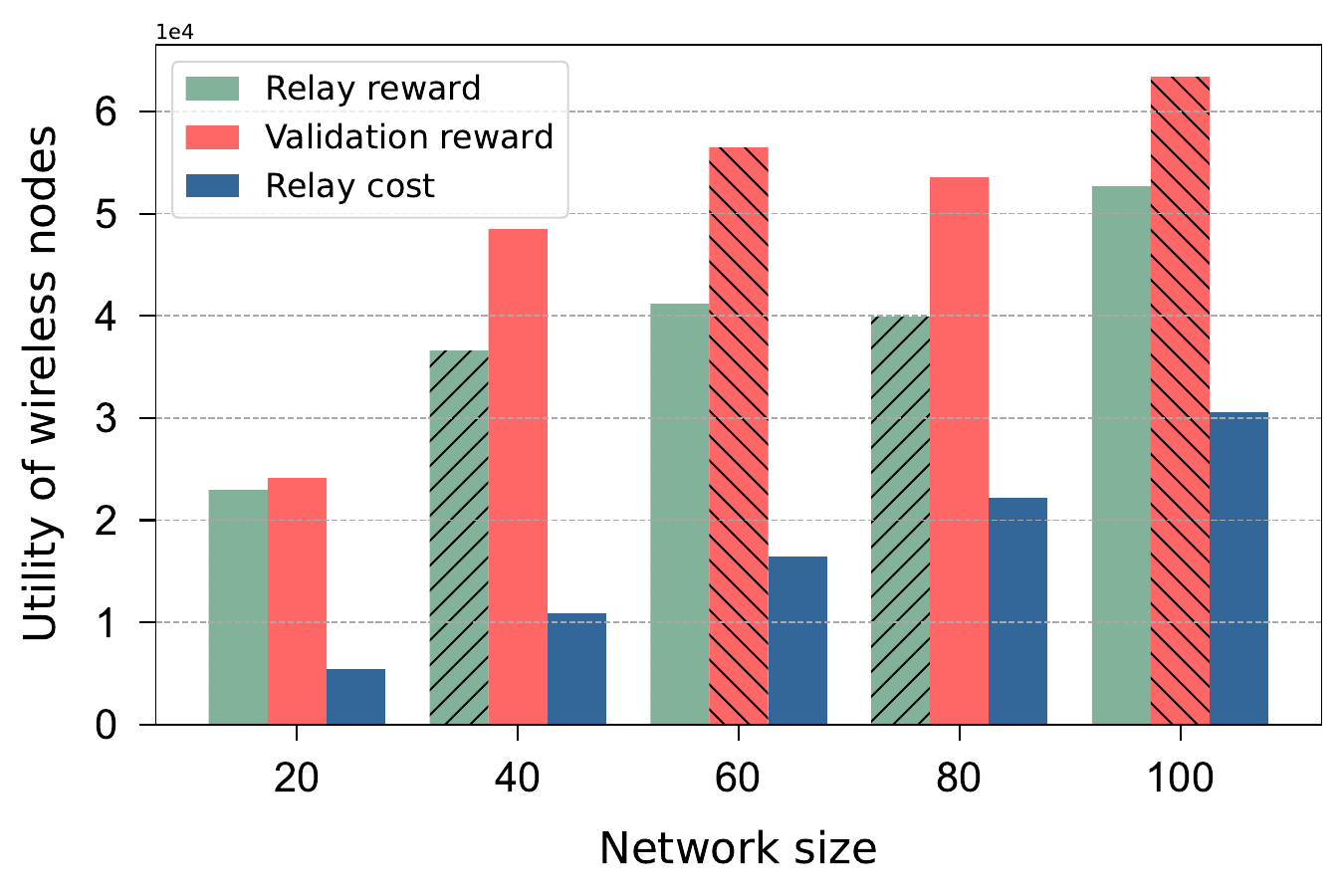}} 
    \ \subfloat[Selfish relay mechanism.]{\includegraphics[scale=0.26]{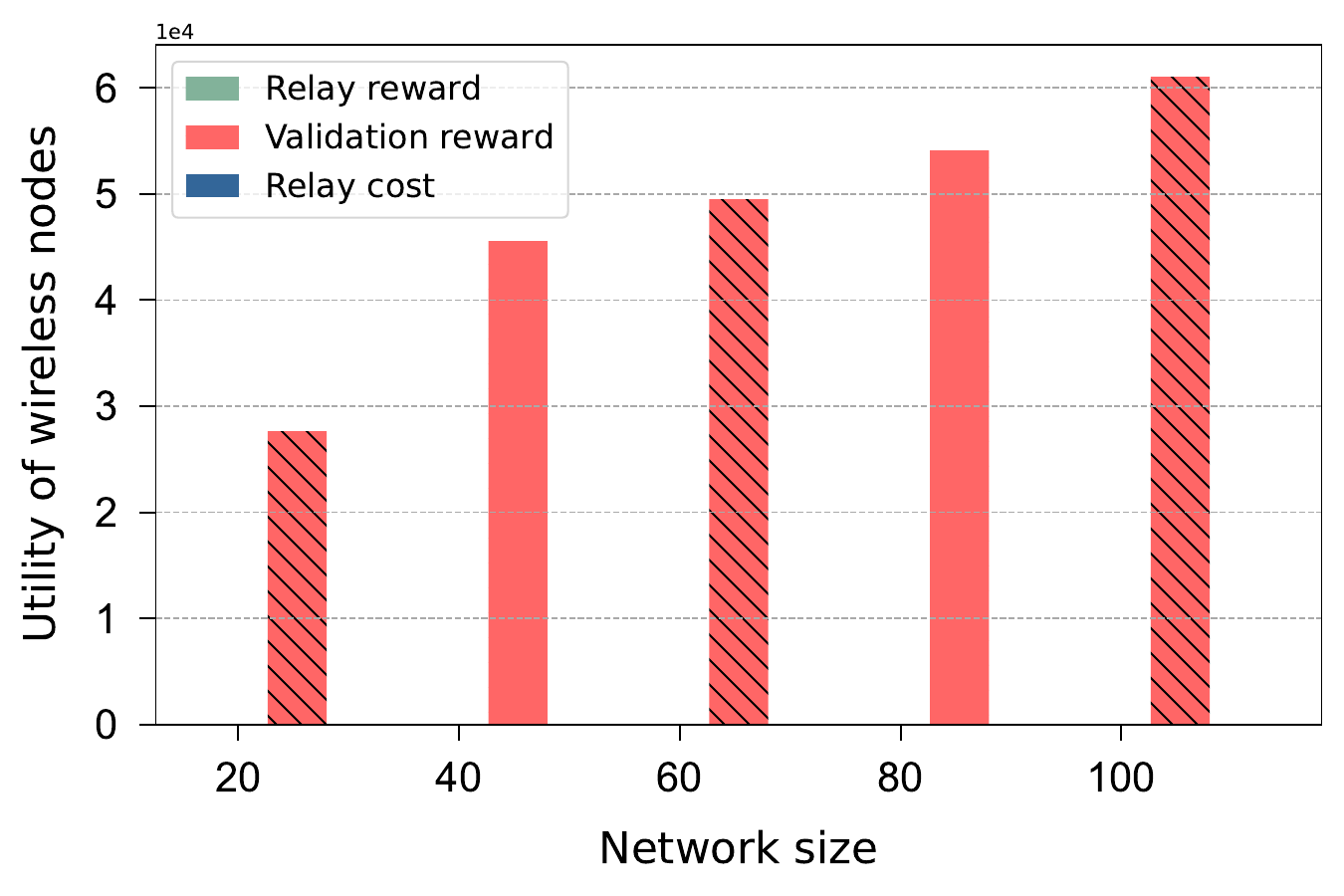}}
    \caption{Transaction fee allocation and relay cost under different incentive mechanisms.}
\end{figure*}

\subsection{Utility of the Proposed Dual Auction Mechanism}

Fig. 7 demonstrate the transaction fee allocation and cost of relaying others' transactions among three mechanisms. The transaction fees are allocated to full nodes under the classical PoW-based and selfish relay mechanisms. We can observe that the relay cost is enormous, and the validation reward is insignificant compared to the relay cost, which means the social welfare is negative. In the selfish relay mechanism, light nodes receive no relay reward and do not relay others' transactions so that they have no relay cost. From Fig. 7(b), both relay and validation behaviors obtain rewards in the dual auction mechanism. The validation reward is higher than the relay reward to ensure full nodes focus on building blocks. Besides, the relay cost is less than the relay reward, which encourages light nodes to relay \emph{Tx\_msg}s.

In Fig. 8, we demonstrate the social welfare of the above three mechanisms. Fig. 8 illustrates that the proposed dual auction mechanism achieves the highest social welfare for the wireless blockchain system. Note that the social welfare declines at network size are $80$. A larger network size brings more transaction fees for nodes, which also consumes higher energy and bandwidth resources. Besides, the TPS is restricted to the block size. Blindly increasing the network size does not constantly improve the TPS of the wireless blockchain network. We will discuss the impact of the block size on TPS in the following sections. The social welfare of the classical PoW-based mechanism is negative since light nodes contribute vast energy and communication resources to relay transactions but receive no rewards. Though the selfish relay mechanism obtains positive social welfare, only minority light nodes profit from the system, which ultimately compromises the long-term operation of the wireless blockchain network. Thus, the dual auction mechanism is appropriate for the wireless blockchain system to encourage light nodes to relay messages and reduce communication consumption in the wireless blockchain network.

\begin{figure}[t]
    \includegraphics[scale=0.35]{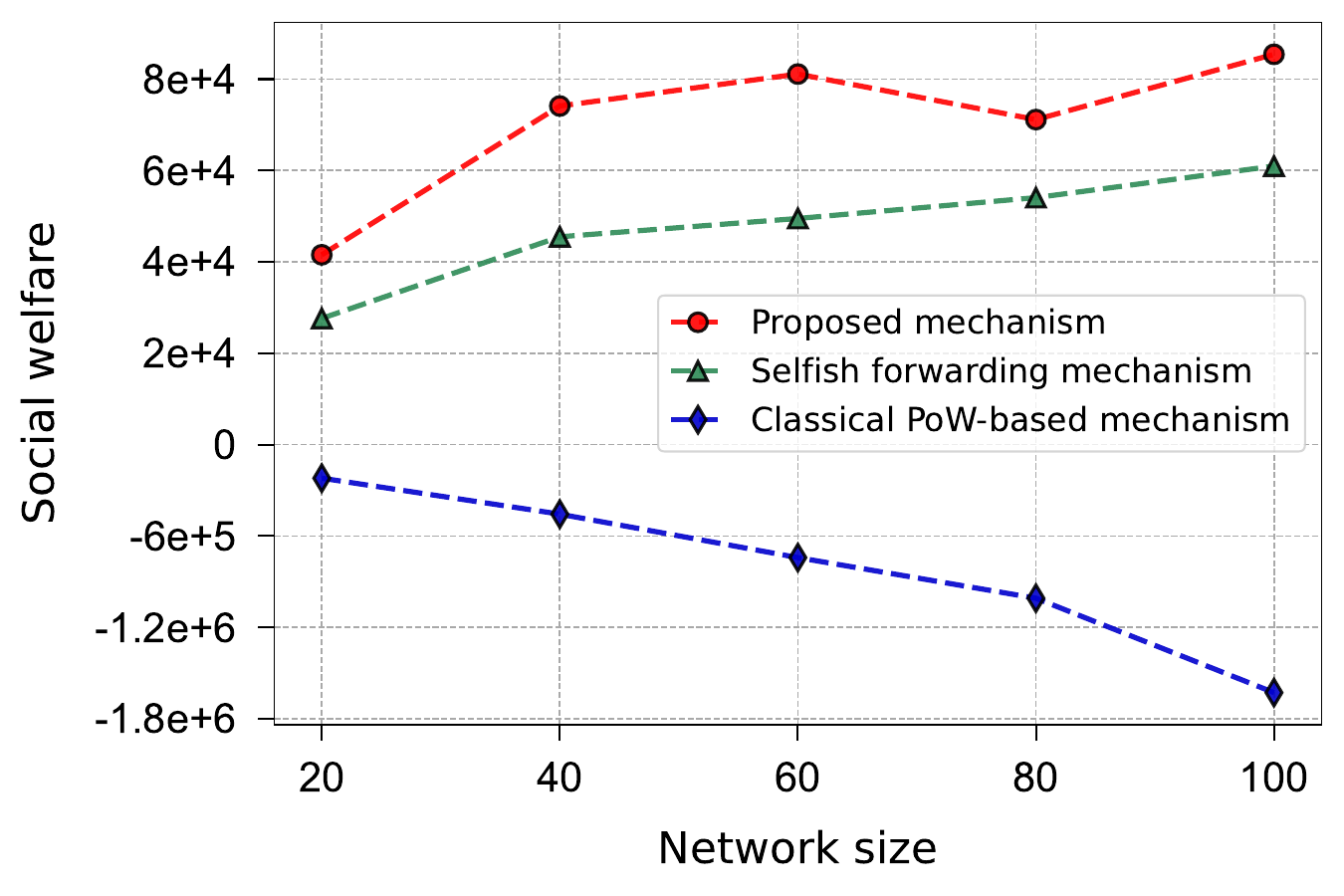}
    \caption{Social welfare under different incentive mechanisms in wireless blockchain networks.}
\end{figure}

\begin{figure}[t]
\includegraphics[scale=0.35]{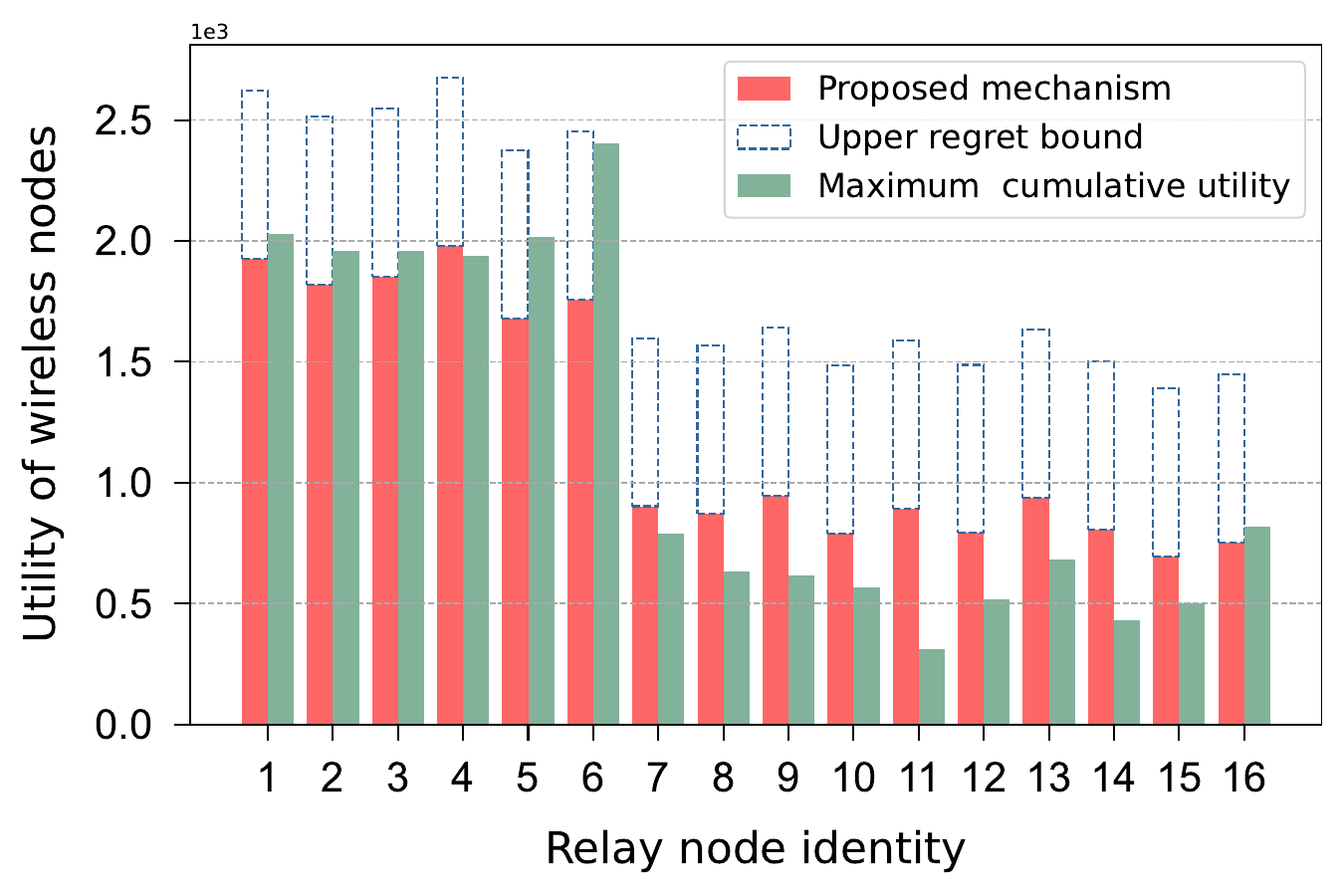}\caption{The difference between the proposed mechanism and maximum maximum cumulative utility.}
\end{figure}

In Fig. 9, we compare the proposed dual auction mechanism with the maximum cumulative utility. As mentioned in Section 5.4, full nodes achieve the optimal utility through validation sub-auction. Thus, we demonstrate the difference between the two mechanisms during the relay sub-auction. We set the network size as $20$ and obtained the total utility of $16$ light nodes in $50$ rounds, respectively. The red bar represents the utility under the dual auction mechanism, the green bar is the maximum cumulative utility, and the dotted bar is the derived upper regret bound. From Fig. 9, we observe that the sum utility of the dual auction mechanism and upper regret bound is always higher than the maximum cumulative utility, proving the theoretical derivation's correctness.

\subsection{Security Analysis of Dual Auction Mechanism}

In this subsection, we analyze the security of the proposed dual auction mechanism under two kinds of adversary models.

Fig. 10 illustrates the Sybil-proof of the designed dual auction mechanism. False identity represents the number of faked relay during a Sybil attack, where $0$ is the honest relay. We conclude that light nodes just relay honestly can achieve the highest utility. Besides, the more false identities are, the lower the utility obtained. On the one hand, our designed reward allocation scheme guarantees that light nodes cannot get extra profit through the Sybil attack. On the other hand, the Sybil attack adds the length of the relay path and damages the interests of other light nodes, reducing the possibility of being relayed. Thus, light nodes are not motivated to launch Sybil attacks under the dual auction mechanism.

\begin{figure}[t]
\includegraphics[scale=0.35]{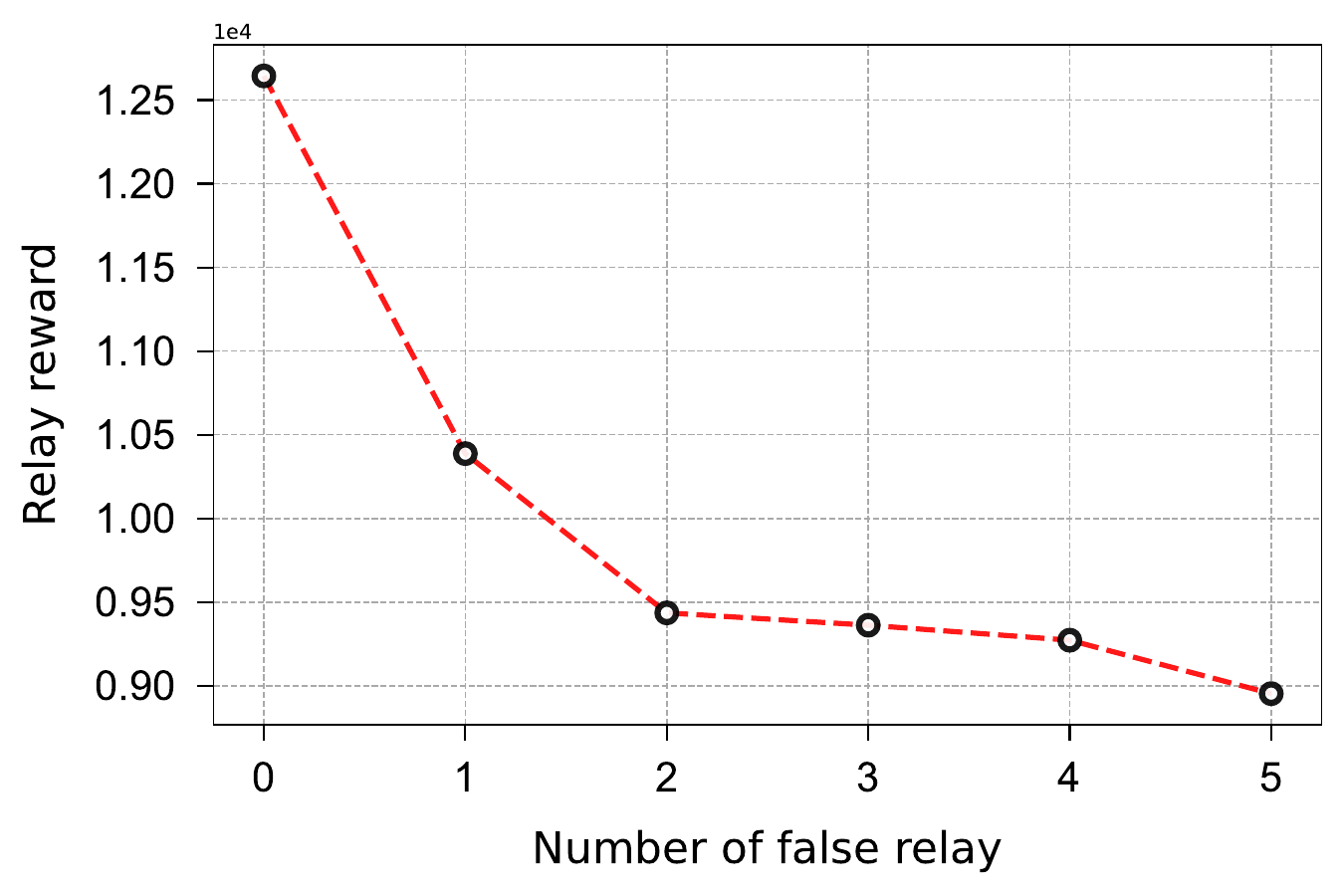}\caption{Sybil-proof of dual auction mechanism.}
\end{figure}

\begin{figure}[t]
\includegraphics[scale=0.35]{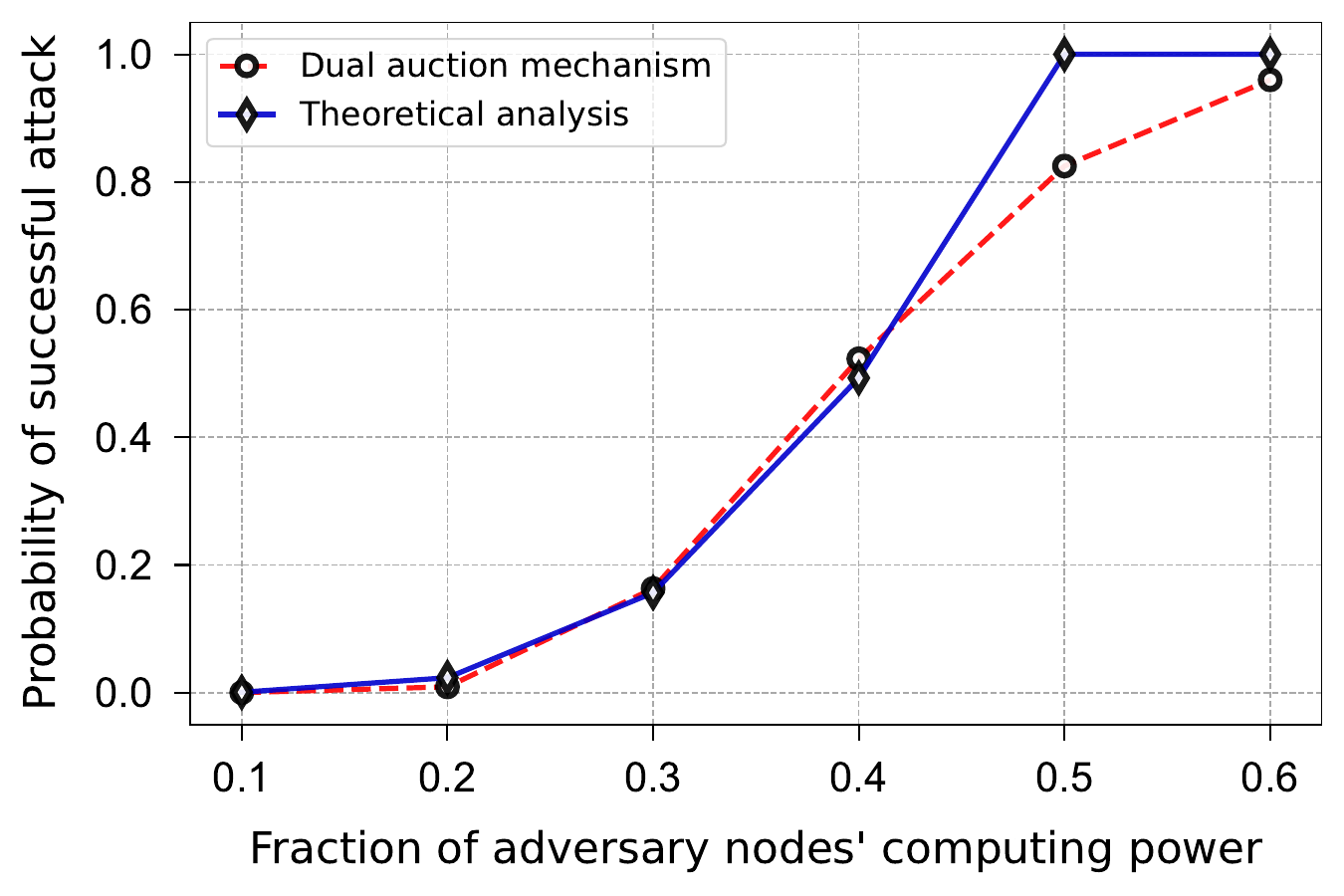}
\caption{Analysis of the double spending attack.}
\end{figure}

\begin{figure}[t]
    \includegraphics[scale=0.35]{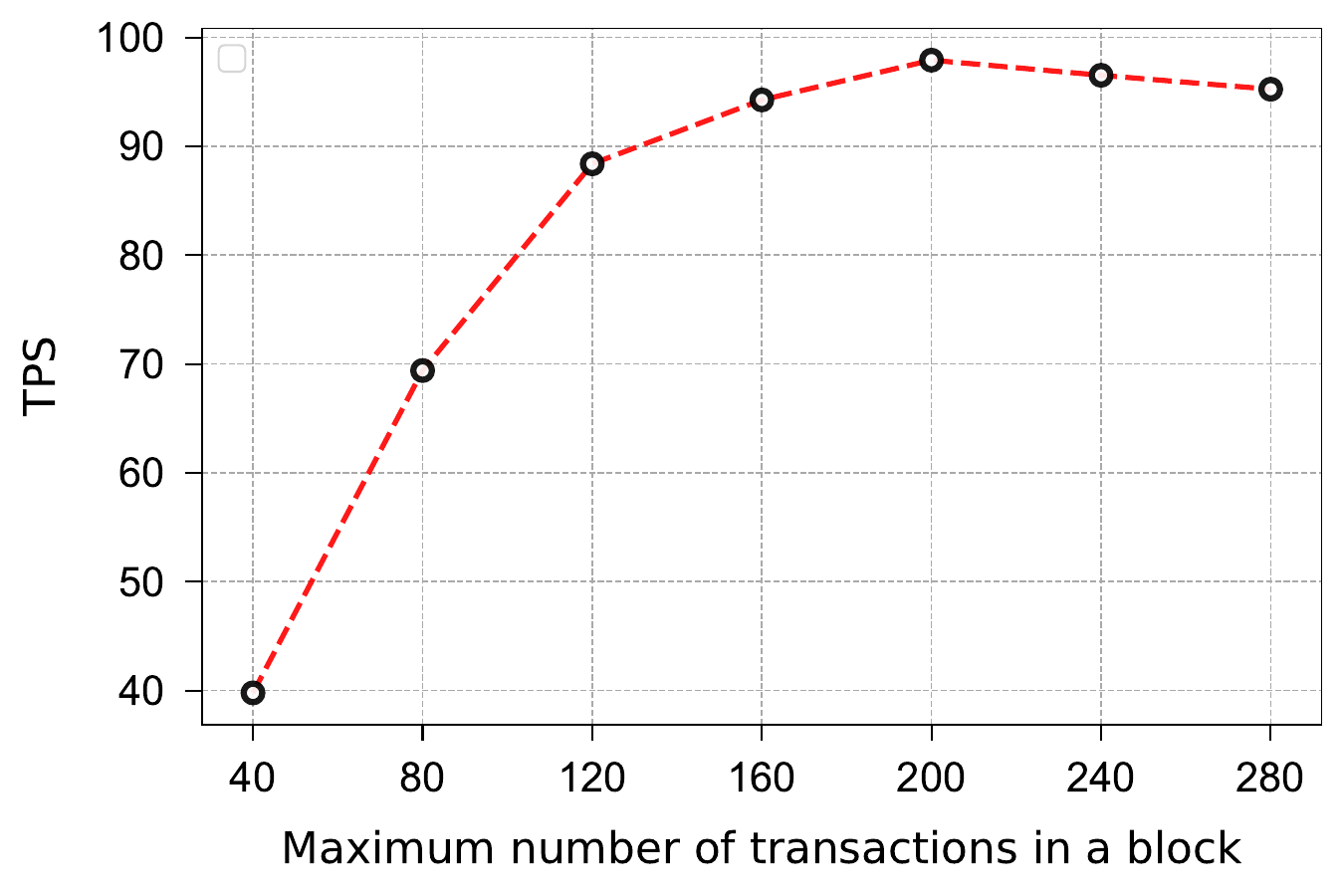}\caption{Impact of the block size.}
\end{figure}

\begin{figure}[t]
\includegraphics[scale=0.35]{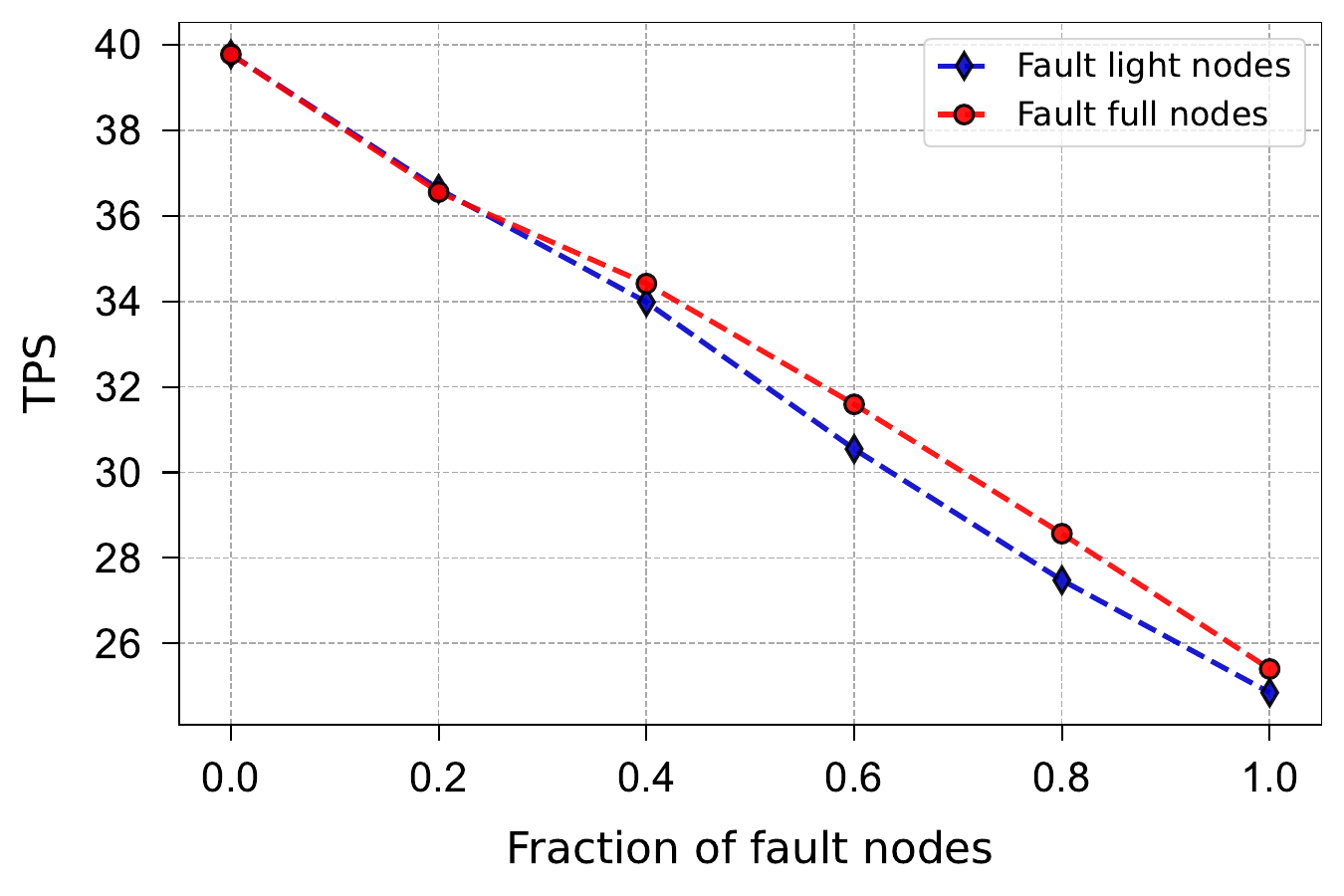}\caption{Impact of the fault nodes.}
\end{figure}

\begin{figure}[th]
\includegraphics[scale=0.35]{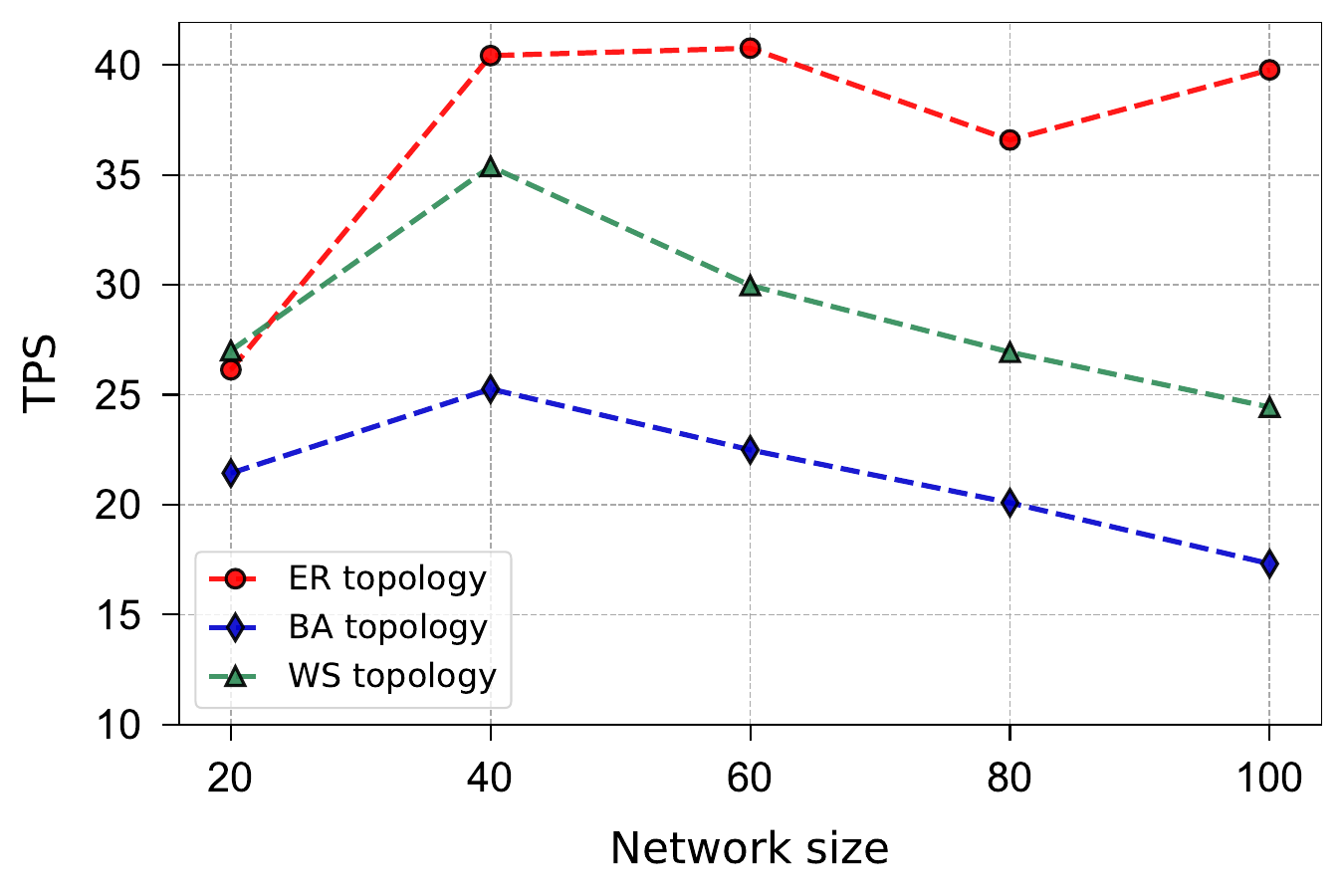}\caption{TPS comparision of wireless blockchain under different complex network models.}
\end{figure}

In Fig. 11, we demonstrate the security threshold of the double spending attack in the proposed dual auction mechanism. By changing the computing power ratio of the malicious full node to the entire network, we obtain the success rate of the double spending attack in $1000$ simulations. From Fig. 11, we observe that the success rate is meager while the fraction of computing power is less than $0.3$. When the fraction of computing power is $0.4$, the malicious full node has a $50$\% success rate of achieving a double spending attack. After the malicious full node occupies more than 50\% computing power of the entire blockchain network, the double spending attack is almost certain to succeed, which is consistent with the analysis of Nakamoto in \cite{rosenfeld2014analysis}. In a nutshell, the dual auction mechanism resists the Sybil attack effectively and has no adverse effect on the security threshold of blockchain meanwhile.

\subsection{Impact of Block Size and Complex Network}

As mentioned before, the TPS is limited by block capacity. In Fig. 12, we analyze block size's impact on the wireless blockchain's TPS. We define the number of \emph{Tx\_msgs} contained in a block as the block size and keep the network size at $100$. As the larger block size increases, the TPS increases and then decreases. At first, the block with a larger block size can contain more \emph{Tx\_msgs}, improving the TPS. However, the transaction generation rate is capped, and a vast block occupies more communication resources, which results in the TPS cannot increase indefinitely as the block size expands. The simulation results show that block size also significantly impacts blockchain performance.

In the wireless blockchain network, nodes usually break down and disconnect from others due to the complex wireless environment. Considering the instability of communication links, we assume there are existing fault nodes in the wireless blockchain network. We define the disruption interval obeys the normal distribution $\mathcal{N}(10s,5s)$, and distribution time follows the normal distribution $\mathcal{N}(0.5s,0.1s)$. There are $80$ light nodes and $20$ full nodes in the wireless blockchain network. We simulate situations about the fault of light and full nodes, respectively. In Fig. 13, we deduce that whatever the situation is, the TPS of the wireless blockchain network decreases with the increase of faulty nodes. Therefore, allocating transaction fees to light nodes for their contributions is reasonable.

Considering the complexity of wireless blockchain networks, we conduct simulations under different complex network topologys to analyze the impact of complex networks on the TPS of wireless blockchain systems. Fig. 14 shows the TPS variation under the ER network, BA network, and WS network. Different complex network topology have a significant influence on TPS. BA and WS networks are more clustered than the ER network. As the network size increases, congestion is more severe under BA and WS networks. Thus, TPS starts to decrease when the network size exceeds 40. The wireless blockchain system achieves the highest TPS in the ER network, which indicates that blockchain fits better with more distributed networks.

\section{Conclusion}

In this paper, aiming at the transaction fee allocation problem, we proposed a dual auction mechanism to incentivize light and full nodes for the wireless blockchain network. The dual auction mechanism includes relay and validation sub-auctions. Relay odes use the relay auction to choose transactions to relay. Full nodes select transactions to construct new blocks through the validation sub-auction. Specifically, we apply the no-regret algorithm to dynamically adjust the relaying probability of light nodes to improve efficiency. Moreover, we derive the upper bound between the proposed mechanism and the optimal scheme, proving that the dual auction is IR, IC, CE, and Sybil-proof. To further prove the dual auction mechanism's effectiveness, we conduct extensive simulations to validate the efficiency and feasibility of the mechanism. 

In future work, we will investigate more elements related to blockchain performance, such as the block size, block interval, and network topology. We will explore a more efficient approach to optimize blockchain performance in the wireless environment by integrating the mechanism design and machine learning.

\section*{Acknowledgment}

This work was supported in part by the National Natural Science Foundation of China under Grants 62101594, U22B2002, and the Natural Science Foundation on Frontier Leading Technology Basic Research Project of Jiangsu under Grant BK20212001.

\bibliographystyle{IEEEtran}
\bibliography{Citation}

\end{document}